\documentclass[10pt]{iopart}
\usepackage{fancybox,graphicx}
\eqnobysec

\usepackage[colorlinks,linkcolor=blue,urlcolor=black,citecolor=blue]{hyperref}

\newcommand{\beq}{\begin{equation}}
\newcommand{\eeq}{\end{equation}}
\newcommand{\beqa}{\begin{eqnarray}}
\newcommand{\eeqa}{\end{eqnarray}}

\newcommand{\abs}[1]{\vert#1\vert}
\newcommand{\coa}{{\rm coa}}
\newcommand{\dd}{{\rm d}}
\newcommand{\eff}{{\rm eff}}
\newcommand{\eps}{{\varepsilon}}
\newcommand{\eq}{{\rm eq}}
\newcommand{\erf}{\mathop{\rm erf}}
\newcommand{\erfc}{\mathop{\rm erfc}}
\newcommand{\frad}[2]{{\displaystyle{\displaystyle#1\over\displaystyle#2}}}
\newcommand{\g}{\gamma}
\newcommand{\half}{\frac{1}{2}}
\newcommand{\ii}{{\rm i}}
\newcommand{\kz}{{\rm KZ}}
\renewcommand{\max}{{\rm max}}
\newcommand{\mean}[1]{\langle#1\rangle}
\renewcommand{\min}{{\rm min}}
\newcommand{\w}[1]{{\widetilde{#1}}}
\newcommand{\F}{\mathrm{F}}
\newcommand{\FL}{\mathrm{FL}}
\renewcommand{\L}{\mathrm{L}}

\begin{document}

\title{The Glauber-Ising chain under low-temperature protocols}

\author{Claude Godr\`eche and Jean-Marc Luck}

\address{Universit\'e Paris-Saclay, CNRS, CEA, Institut de Physique Th\'eorique,
91191~Gif-sur-Yvette, France}

\begin{abstract}
This work is devoted to an in-depth analysis of arbitrary temperature protocols
applied to the ferromagnetic Glauber-Ising chain
launched from a disordered initial state and evolving in the low-temperature scaling regime.
We focus our study on the density of domain walls and the reduced susceptibility.
Both the inverse of the former observable and the latter one provide two
independent measures of the typical size
of the growing ferromagnetic domains.
Their product is thus a dimensionless form factor
characterising the pattern of growing ordered domains
and providing a measure of the distance of the system to thermal equilibrium.
We apply this framework to a variety of protocols: everlasting slow quenches,
where temperature decreases continuously to zero in the
limit of infinitely long times, slow quenches of finite duration,
where temperature reaches zero at some long but finite quenching time,
time-periodic protocols with weak and strong modulations,
and the two-temperature protocol
leading to the memory effect found by Kovacs.
\end{abstract}

\address{\today}

\eads{\mailto{claude.godreche@ipht.fr},\mailto{jean-marc.luck@ipht.fr}}

\maketitle

\section{Introduction}
\label{intro}

This work is at the confluence of two streams of ideas.
On the one hand,
in the wake of the study by Glauber of the single-spin-flip dynamics of the
Ising chain~\cite{glauber}, there has been a continuous thread of research
aimed at extending the exactly solvable Glauber dynamics to arbitrary
temperature-change protocols~\cite{reiss,schilling,brey}.
In the original scheme considered by Glauber,
the system evolves at some fixed temperature~$T$, starting from a random initial condition,
which is equivalent to saying that the system is instantaneously quenched
from infinite temperature down to the finite temperature~$T$.
The studies made in~\cite{reiss,schilling,brey} extend this simple protocol to more general ones,
such as slower quenches or diverse cooling and heating scenarios.

A second stream of ideas concerns the so-called Kibble-Zurek mechanism~\cite{kibble,zurek}
which takes place when a system is progressively quenched at some finite rate
through a continuous phase transition.
The pioneering work by Kibble was motivated by the cooling of our Universe
in its primordial history.
It was then adapted by Zurek to condensed matter systems,
and resulted in an approximate scheme to estimate
the density of residual defects,
and more generally the residual excess energy at the end of the quench,
as a function of the quenching rate.
The Kibble-Zurek mechanism has been investigated in the Ising
chain~\cite{krapiv,jeong,priyanka,mayo}
and in the Ising model and other ferromagnets in higher dimensions~\cite{biroli,ceg}.

Here we revisit these themes with a new eye.
This work introduces two main novelties.
First, we provide a considerable simplification of earlier treatments of the
dynamics of the Glauber-Ising chain,
by focussing our attention on the low-temperature scaling regime.
Second, we consider in parallel two concurrent observables,
both characterising the typical growing length of ferromagnetic domains,
namely the density~$\rho(t)$ of domain walls on the one hand, and the reduced
susceptibility~$\chi(t)$ on the other.
Their dimensionless product $\Pi(t)$ is a form factor
characterising the whole pattern of growing ordered domains,
providing thus a measure of the distance of the system to thermal equilibrium
(see~(\ref{rhodef})--(\ref{pidef})).

Consider the ferromagnetic Ising chain defined by the Hamiltonian
\beq
{\cal H}=-J\sum_n\sigma_n\sigma_{n+1}.
\eeq
Henceforth we choose dimensionless energy units by setting $J=1$.
In line with earlier studies~\cite{reiss,schilling,brey,krapiv,jeong,priyanka,mayo},
we consider the situation where the temperature~$T(t)$ depends on time~$t$
according to some given protocol.
We denote the inverse temperature by $\beta(t)=1/T(t)$.

The chain evolves under the single-spin-flip heat-bath or Glauber dynamics
where, in the thermodynamic limit of an infinitely long chain,
each spin is updated at Poissonian times,
according to the rule~\cite{glauber}
\beq
\sigma_n(t)\to-\sigma_n(t),
\eeq
with rate
\beqa
\omega_n(t)
&=&\half\Bigl(1-\sigma_n\tanh\bigl(\beta(t)(\sigma_{n-1}(t)+\sigma_{n+1}(t))\bigr)\Bigr)
\nonumber\\
&=&\half\bigl(1-\g(t)\sigma_n(t)(\sigma_{n-1}(t)+\sigma_{n+1}(t))\bigr),
\label{omegadef}
\eeqa
where
\beq
\g(t)=\tanh 2\beta(t).
\eeq
The linearization of the tanh function involved in going from the first
to the second line of~(\ref{omegadef}) is the key point which makes Glauber dynamics solvable.

Throughout the following we consider the dynamics of the Ising chain
under an arbitrary temperature protocol $\beta(t)$,
starting from the translationally invariant disordered initial state
where all spin configurations are equally probable.
We concentrate on the equal-time two-point correlation function
\beq
C_n(t)=\mean{\sigma_n(t)\sigma_0(t)},
\label{cndef}
\eeq
where $\mean{\cdot}$ denotes an average over the disordered initial state
of the system and over its stochastic history.
We pay special attention to the low-temperature dynamical scaling regime
which takes place in the vicinity of the critical point $T_c=0$,
focussing our attention on the following observables:

\begin{itemize}

\item
the density of domain walls
\beq
\rho(t)=\half(1-C_1(t))=\half(1+E(t)),
\label{rhodef}
\eeq
where $E(t)=-C_1(t)$ is the energy density of the system,

\item
the reduced susceptibility
\beq
\chi(t)=\sum_nC_n(t),
\label{chidef}
\eeq

\item
the form factor
\beq
\Pi(t)=\rho(t)\chi(t).
\label{pidef}
\eeq

\end{itemize}

Both $1/\rho(t)$ and $\chi(t)$ provide measures of the
typical size of the growing ferromagnetic domains,
so that the ratio of these lengths, namely the form factor~$\Pi(t)$,
provides a dimensionless parameter characterising the state of the system.
The non-triviality of the form factor $\Pi(t)$
mirrors the property that domain lengths
are neither mutually independent nor exponentially distributed~\cite{dz},
except at thermal equilibrium, where we have $\Pi_\eq=1$.

This paper is structured as follows.
In section~\ref{fin} we revisit the essentials of the exactly solvable
Glauber-Ising chain under an arbitrary temperature protocol,
obtaining thus in a rather direct way the exact expressions~(\ref{rhores}) and~(\ref{chires})
for the main observables of interest $\rho(t)$ and $\chi(t)$.
We then investigate at depth in section~\ref{low}
the dynamics of the Glauber-Ising chain under an arbitrary protocol
in the low-temperature scaling regime in the vicinity of the critical point $T_c=0$.
Equations~(\ref{rhopro}) and~(\ref{chipro}) are our key outcomes,
providing expressions for $\rho(t)$ and $\chi(t)$
in the case of an arbitrary protocol.
These scaling results are more amenable to analytical treatment than
their exact finite-temperature analogues.

Subsequent sections are devoted to applying the above formalism
to diverse low-temperature protocols.
In section~\ref{ever} we consider an everlasting slow quench, where
temperature decreases continuously from infinity to zero at infinitely long times.
The scaling properties of the system at long times
only depend on the tail of the quenching process.
The system either obeys an effective finite-temperature near-equilibrium
dynamics (adiabatic approximation),
or an effective zero-temperature non-equilibrium dynamics (coarsening).
This qualitative picture is in full agreement with a heuristic reasoning
germane to the Kibble-Zurek scaling theory.
It is complemented by quantitative analytical predictions.
In section~\ref{slofin} we consider, as in~\cite{brey,krapiv,jeong,priyanka,mayo},
a slow quench of finite duration,
where temperature decreases continuously and reaches zero
at some long but finite quenching time $\tau$.
We consider the system at the quenching time ($t=\tau$)
and in the late-time regime ($t\gg\tau$) in turn.
In the former case,
scaling properties again only depend on the tail of the quenching process.
Our results corroborate the power laws
$1/\rho(\tau)\sim\chi(\tau)\sim\tau^\lambda$,
in qualitative accord with Kibble-Zurek scaling theory.
We derive the exact amplitudes of the density of defects $\rho(\tau)$ and of the
susceptibility $\chi(\tau)$ and the ensuing universal value of the form factor $\Pi(\tau)$.
Section~\ref{per} is devoted to time-periodic protocols.
We investigate in detail the case of a harmonic periodic driving
with arbitrary frequency.
We successively investigate the linear-response regime of a weak modulation,
and the case of a critical modulation, where temperature vanishes once per period.
In section~\ref{ko} we provide a quantitative analysis of the Kovacs effect
throughout the low-temperature scaling regime.
A brief discussion of our results is given in section~\ref{disc}.
We relegate some technical material to two appendices.

\section{Exact results at finite temperature}
\label{fin}

\subsection{Generalities on finite-temperature Glauber dynamics}
\label{fingen}

In this section we recall how the two-point function $C_n(t)$
can be determined exactly for the Ising chain with Glauber dynamics
with an arbitrary temperature protocol $\beta(t)$.
We shall thus recover by simpler means
the outcomes of several earlier works~\cite{reiss,schilling,brey}.
This section also sets the stage for the extensive analysis
of various specific protocols in the low-temperature scaling regime,
to be performed in the next sections.

The starting point of the analysis dates back to Glauber~\cite{glauber}.
The specific form~(\ref{omegadef}) of the flipping rate
implies that the two-point function $C_n(t)$ obeys the linear differential equations
\beq
\frac{\dd C_n(t)}{\dd t}=-2C_n(t)+\g(t)(C_{n-1}(t)+C_{n+1}(t))\qquad(n\ne0),
\label{cindot}
\eeq
together with the normalization condition
\beq
C_0(t)=1.
\label{cnor}
\eeq

The evolution equations~(\ref{cindot}) are linear, and so they are amenable to an exact solution.
The only delicate part consists in imposing the condition~(\ref{cnor}).
Several approaches have been used for that purpose in the past.
In line with our previous work~\cite{gl2000},
we choose to complement the coupled equations~(\ref{cindot})
by their counterpart for $n=0$, involving a time-dependent source term $w(t)$:
\beq
\frac{\dd C_n(t)}{\dd t}=-2C_n(t)+\g(t)(C_{n-1}(t)+C_{n+1}(t))+w(t)\delta_{n0}.
\label{cdot}
\eeq
The source term $w(t)$ is to be determined from the condition
that the condition~(\ref{cnor}) holds at all times.
It is related to the density of domain walls as
\beq
w(t)=2-2\g(t)C_1(t)=2-2\g(t)+4\g(t)\rho(t).
\label{wrho}
\eeq
This expression is obtained by equating to zero
the right-hand-side of~(\ref{cdot}) for $n=0$.

The discrete Fourier transform $C^\F(q,t)$ (see~\ref{appfl} for conventions),
also known as the time-dependent structure factor, obeys
\beq
\frac{\dd C^\F(q,t)}{\dd t}=-2(1-\g(t)\cos q)C^\F(q,t)+w(t).
\label{cqdot}
\eeq
The condition~(\ref{cnor}) reads
\beq
\int_0^{2\pi}\frac{\dd q}{2\pi} C^\F(q,t)=1,
\label{cqnor}
\eeq
whereas the assumption of a disordered initial state translates to
\beq
C_n(0)=\delta_{n0},
\eeq
i.e.,
\beq
C^\F(q,0)=1.
\label{cqinit}
\eeq

\subsection{The case of a quench to a constant temperature}
\label{fincst}

Before considering an arbitrary temperature protocol,
we first review the case where the Glauber-Ising chain
relaxes from a disordered initial state at constant temperature (see~\cite{gl2000}).

In this case, $\g$ is constant, and so~(\ref{cqdot}),
with initial condition~(\ref{cqinit}),
reads in Fourier-Laplace space (see~\ref{appfl} for conventions)
\beq
pC^\FL(q,p)=-2(1-\g\cos q)C^\FL(q,p)+w^\L(p)+1.
\eeq
We readily obtain
\beq
C^\FL(q,p)=\frac{w^\L(p)+1}{p+2-2\g\cos q}.
\eeq
The condition~(\ref{cqnor}) translates to
\beq
\frac{1}{p}=\int_0^{2\pi}\frac{\dd
q}{2\pi}C^\FL(q,p)=\frac{w^\L(p)+1}{\sqrt{(p+2)^2-4\g^2}},
\label{walg}
\eeq
hence
\beq
w^\L(p)=\frac{\sqrt{(p+2)^2-4\g^2}}{p}-1
\eeq
and
\beq
C^\FL(q,p)=\frac{\sqrt{(p+2)^2-4\g^2}}{p(p+2-2\g\cos q)}.
\label{cfl}
\eeq
This closed-form result contains all the relevant information
on the two-point correlations during a quench to a constant temperature.

In the limit of infinitely long times, the system reaches thermal equilibrium.
The equilibrium structure factor reads
\beq
C^\F_\eq(q)=\lim_{p\to0}pC^\FL(q,p)=\frac{\sqrt{1-\g^2}}{1-\g\cos q}.
\eeq
This expression can be recast as
\beq
C^\F_\eq(q)=\frac{\sinh\mu}{\cosh\mu-\cos q},
\label{ce}
\eeq
with
\beq
\g=\tanh 2\beta=\frac{1}{\cosh\mu}.
\label{gammares}
\eeq
The equilibrium correlations therefore decay exponentially with distance, as
\beq
C_{n,\eq}=\e^{-\mu\abs{n}}.
\label{ceqres}
\eeq
We thus recover the expression of the equilibrium correlation length
$\xi_\eq$~\cite{baxter}:
\beq
\e^{-1/\xi_\eq}=\e^{-\mu}=\tanh\beta.
\label{mures}
\eeq
We obtain in particular the following expressions:
\beqa
\rho_\eq&=&\half(1-\e^{-\mu})=\frac{1}{\e^{2\beta}+1},
\nonumber\\
\chi_\eq&=&\coth\frac{\mu}{2}=\frac{1+\e^{-\mu}}{1-\e^{-\mu}}=\e^{2\beta},
\nonumber\\
\Pi_\eq&=&\half(1+\e^{-\mu})=\frac{\e^{2\beta}}{\e^{2\beta}+1}.
\label{pieq}
\eeqa

At any finite temperature,
the system converges exponentially fast to thermal equilibrium.
The relaxation time characterizing this convergence depends a priori on the observable.
As far as equal-time correlations are concerned,
the nearest pole of~(\ref{cfl}) takes place for $q=0$
and yields the relaxation time
\beq
\tau_\eq=\frac{1}{2(1-\g)}=\frac{\cosh\mu}{2(\cosh\mu-1)}=\frac{\e^{4\beta}+1}{4}.
\label{taures}
\eeq

\subsection{The case of an arbitrary temperature protocol}
\label{fingal}

In the case of an arbitrary temperature protocol specified by some given
$\g(t)$,
the solution to~(\ref{cqdot}), with initial condition~(\ref{cqinit}), reads
\beq
C^\F(q,t)=\e^{-2t+2g(t)\cos q}
+\int_0^t\e^{-2(t-s)+2(g(t)-g(s))\cos q}\,w(s)\dd s,
\label{csol}
\eeq
with
\beq
g(t)=\int_0^t\g(s)\dd s.
\label{gdef}
\eeq
The condition~(\ref{cqnor}) translates to
\beq
\e^{-2t}I_0(2g(t))+\int_0^t\e^{-2(t-s)}I_0(2g(t)-2g(s))\,w(s)\dd s=1,
\label{weq}
\eeq
where $I_0$ is the modified Bessel function (see~\ref{appb}).

In order to extract the source term $w(t)$ from the integral equation~(\ref{weq}),
we consider $g(t)$ as a new temporal variable,
denote the inverse function as $t(g)$,
and introduce the notations
\beq
\w\g(g)=\g(t(g)),\qquad\w w(g)=w(t(g)),
\label{wnots}
\eeq
so that
\beq
\frac{\dd t}{\dd g}=\frac{1}{\w\g(g)}.
\label{dtdg}
\eeq
Equation~(\ref{weq}) then reads
\beq
I_0(2g)+\int_0^g\e^{2t(h)}I_0(2g-2h)\frac{\w w(h)}{\w\g(h)}\dd h=\e^{2t(g)}.
\label{wint}
\eeq
Introducing the combination
\beq
\Lambda(g)=\delta(g)+\frac{\w w(g)}{\w\g(g)},
\label{lamdef}
\eeq
where $\delta(g)$ is Dirac's delta function,~(\ref{wint}) translates to
\beq
\int_0^g\Lambda(h)\e^{2t(h)}I_0(2g-2h)\dd h=\e^{2t(g)}.
\label{wcon}
\eeq
The integral has the form of a convolution,
so that it is natural to use Laplace transforms.
Defining
\beqa
L_0(p)&=&\int_0^\infty\e^{2t(g)-pg}\dd g,
\nonumber\\
L_1(p)&=&\int_0^\infty\Lambda(g)\,\e^{2t(g)-pg}\dd g,
\nonumber\\
L_2(p)&=&2\int_0^\infty\left(\frac{1}{\w\g(g)}-1\right)\e^{2t(g)-pg}\dd g,
\label{L0L1L2}
\eeqa
the integral equation~(\ref{wcon}) becomes
(see entry (1) of Table~\ref{ilts})
\beq
\frac{L_1(p)}{\sqrt{p^2-4}}=L_0(p),
\label{L1L0}
\eeq
whereas~(\ref{dtdg}) implies
\beq
L_2(p)=(p-2)L_0(p)-1.
\eeq
Eliminating $L_0(p)$ between both above expressions, we are left with
\beq
L_1(p)=\sqrt\frac{p+2}{p-2}\;(L_2(p)+1).
\label{L1L2}
\eeq

The inverse Laplace transform of the first factor in the right-hand side
is given by entry~(2) of Table~\ref{ilts}.
Inserting the latter expression into~(\ref{L1L2}), using the
definitions~(\ref{L0L1L2}),
and translating back to the original time $t$,
we obtain the following expression for the source term
corresponding to an arbitrary temperature protocol:
\beqa
w(t)
&=&2(1-\g(t))+2\g(t)\e^{-2t}(I_0(2g(t))+I_1(2g(t)))
\nonumber\\
&+&4\g(t)\int_0^t(1-\g(s))\e^{-2(t-s)}
\nonumber\\
&\times&(I_0(2g(t)-2g(s))+I_1(2g(t)-2g(s)))\,\dd s,
\label{wres}
\eeqa
where $I_0$ and $I_1$ are modified Bessel functions (see~\ref{appb}).
We recall that $g(t)$ has been defined in~(\ref{gdef}).
An expression for the density of domain walls
can be obtained by using~(\ref{wrho}):
\beqa
\rho(t)
&=&\half\e^{-2t}(I_0(2g(t))+I_1(2g(t)))
\nonumber\\
&+&\int_0^t(1-\g(s))\e^{-2(t-s)}
\nonumber\\
&\times&(I_0(2g(t)-2g(s))+I_1(2g(t)-2g(s)))\,\dd s.
\label{rhores}
\eeqa

The susceptibility $\chi(t)$ can be investigated along the same lines.
Equation~(\ref{csol}) yields
\beq
\chi(t)=C^\F(0,t)=\e^{-2t+2g(t)}
+\int_0^t\e^{-2(t-s)+2(g(t)-g(s))}\,w(s)\dd s.
\label{chisol}
\eeq
Using notations similar to~(\ref{wnots}),
the relation~(\ref{chisol}) can be recast as
\beq
\w\chi(g)\e^{2t(g)-2g}=\int_0^g\Lambda(h)\e^{2t(h)-2h}\dd h,
\eeq
implying that the Laplace transform
\beq
L_3(p)=\int_0^\infty\w\chi(g)\,\e^{2t(g)-(p+2)g}\dd g
\eeq
reads (see~(\ref{L0L1L2}),~(\ref{L1L0}))
\beq
L_3(p)=\frac{L_1(p+2)}{p}=\sqrt\frac{p+4}{p}\,L_0(p+2).
\eeq
The inverse Laplace transform of the first factor in the rightmost side
is again given by entry~(2) of Table~\ref{ilts}.
Pursuing along the lines of the derivation of~(\ref{wres}),
we obtain
\beqa
\chi(t)
&=&1+2\int_0^t\g(s)\e^{-2(t-s)}
\nonumber\\
&\times&(I_0(2g(t)-2g(s))+I_1(2g(t)-2g(s)))\,\dd s.
\label{chires}
\eeqa

The formulas~(\ref{rhores}) and~(\ref{chires}) have essentially the same
structure
as the key outcomes of several earlier works~\cite{reiss,schilling,brey}.
They do not easily lend themselves to a thorough analysis in
specific examples of temperature protocols.
In section~\ref{low} we show how the analysis simplifies
in the low-temperature scaling regime in the vicinity of the critical point $T_c=0$.

Note that for generic values of the momentum $q$,
an expression of the structure factor~$C^\F(q,t)$ involving a double time integral
can be obtained by inserting~(\ref{wres}) into~(\ref{csol}).
At variance with~(\ref{chires}),
the resulting expression cannot be reduced to a single integral.

\section{Low-temperature scaling regime}
\label{low}

\subsection{Characteristic length and time scales}
\label{lowchar}

It is well-known that the ferromagnetic Ising chain
has no long-range order at finite temperature,
so that its critical temperature is $T_c=0$~\cite{baxter}.
Its equilibrium properties
exhibit a low-temperature scaling regime in the vicinity of the critical point.
The characteristic length scale is the correlation length~$\xi_\eq$,
which gives a measure of the typical size of ferromagnetic domains at thermal equilibrium.
This length diverges exponentially fast at low temperature, according to
\beq
\xi_\eq=\frac{1}{\mu}\approx\frac{\e^{2\beta}}{2}.
\eeq
It is convenient to express all equilibrium quantities
in terms of the inverse equilibrium correlation length $\mu$.
We have in particular (see~(\ref{ce}))
\beq
C^\F_\eq(q)\approx\frac{2\mu}{q^2+\mu^2}
\label{cfeq}
\eeq
and
\beq
\rho_\eq\approx\frac{\mu}{2},\qquad
\chi_\eq\approx\frac{2}{\mu},
\label{rceq}
\eeq
so that the form factor is asymptotically equal to unity:
\beq
\Pi_\eq=1.
\label{pie}
\eeq
The relaxation time diverges as (see~(\ref{taures}))
\beq
\tau_\eq\approx\frac{1}{\mu^2}.
\label{taueq}
\eeq
The scaling law $\tau_\eq\sim\xi_\eq^2$ evidences the dynamical exponent $z=2$,
in accord with the diffusive motion of domain walls under zero-temperature
Glauber dynamics.

It will be shown below that the above scaling picture extends to
the dynamics of the model subjected to an arbitrary low-temperature protocol.

\subsection{Generalities on Glauber dynamics in the low-temperature scaling
regime}
\label{lowgen}

The aim of this section is to show how the approach of Section~\ref{fin}
simplifies in the low-temperature scaling regime,
where the characteristic length and time scales diverge.
We find it easier and more didactic to resume the analysis from the very
beginning,
rather than looking for the scaling behaviour of various outcomes
of Section~\ref{fin}.

Throughout the following,
we parametrize the temperature protocol by means of the instantaneous inverse
correlation length,
\beq
\mu(t)=-\ln\tanh\beta(t),
\eeq
which becomes exponentially small at low temperatures,
where we have
\beq
\mu(t)\approx2\,\e^{-2\beta(t)}
\eeq
and
\beq
\g(t)\approx1-\frac{\mu^2(t)}{2}.
\eeq
The relevant momentum scale $q$ is accordingly small
in the low-temperature scaling regime,
so that the discrete Fourier transform entering~(\ref{cqdot})
becomes a continuous one (see~\ref{appfl} for conventions).
Equation~(\ref{cqdot}) thus reads
\beq
\frac{\dd C^\F(q,t)}{\dd t}\approx-(q^2+\mu^2(t))C^\F(q,t)+w(t),
\label{ccqdot}
\eeq
whereas the relation~(\ref{wrho}) between the source term and the density of
domain walls
boils down to
\beq
w(t)\approx4\rho(t).
\label{cwrho}
\eeq
The condition~(\ref{cnor}) reads
\beq
\int_{-\infty}^{+\infty}\frac{\dd q}{2\pi} C^\F(q,t)=1,
\label{ccqnor}
\eeq
whereas the assumption of a disordered initial state still translates to
\beq
C^\F(q,0)=1.
\label{ccqinit}
\eeq

\subsection{The case of a quench to a constant low temperature}
\label{lowcst}

Before considering an arbitrary temperature protocol, let us,
as in section~\ref{fincst}, consider first the case of a quench from a disordered
initial state
to a constant low temperature in the scaling regime.

In this case, $\mu(t)=\mu$ is constant, and so~(\ref{ccqdot}),
with initial condition~(\ref{ccqinit}),
reads in continuous Fourier-Laplace space (see~\ref{appfl} for conventions)
\beq
pC^\FL(q,p)\approx-(q^2+\mu^2)C^\FL(q,p)+w^\L(p)+1.
\eeq
This equation can be solved readily, yielding
\beq
C^\FL(q,p)\approx\frac{w^\L(p)+1}{p+q^2+\mu^2}.
\eeq
The condition~(\ref{ccqnor}) translates to
\beq
\frac{1}{p}=\int_{-\infty}^{+\infty}\frac{\dd q}{2\pi}C^\FL(q,p)
\approx\frac{w^\L(p)+1}{2\sqrt{p+\mu^2}},
\label{cwalg}
\eeq
so that
\beq
w^\L(p)\approx\frac{2\sqrt{p+\mu^2}}{p}
\label{cwres}
\eeq
is much larger than unity in the scaling regime.
We thus obtain
\beq
C^\FL(q,p)\approx\frac{2\sqrt{p+\mu^2}}{p(p+q^2+\mu^2)}.
\label{ccfl}
\eeq
This formula is the continuum analogue of~(\ref{cfl}),
and reproduces~(\ref{cfeq}) in the static limit.

Inverting first the Fourier transform involved in~(\ref{ccfl}),
and then the Laplace transform (see entry~(3) of Table~\ref{ilts}),
we obtain the scaling form of the correlation function:
\beq
C_n(t)\approx\half\left(\e^{\mu\abs{n}}\erfc\frac{\abs{n}+2\mu t}{2\sqrt{t}}
+\e^{-\mu\abs{n}}\erfc\frac{\abs{n}-2\mu t}{2\sqrt{t}}\right),
\eeq
which is a function of the scaling variables $n/\sqrt{t}$ and
$\mu^2t\approx t/\tau_\eq$, as highlighted in~\cite{gl2000}.

Equations~(\ref{cwres}) and~(\ref{ccfl}) yield in particular
\beqa
\rho^\L(p)\approx\frac{w^\L(p)}{4}\approx\frac{\sqrt{p+\mu^2}}{2p},
\nonumber\\
\chi^\L(p)=C^\FL(0,p)\approx\frac{2}{p\sqrt{p+\mu^2}}.
\eeqa
The Laplace transforms can be inverted explicitly
(see entries~(4) and (5) of Table~\ref{ilts}).
We thus obtain
\beqa
\rho(t)&\approx&\half\left(\frac{\e^{-\mu^2t}}{\sqrt{\pi
t}}+\mu\erf(\mu\sqrt{t})\right),
\label{rhosca}
\\
\chi(t)&\approx&\frac{2}{\mu}\,\erf(\mu\sqrt{t}),
\label{chisca}
\\
\Pi(t)&\approx&\erf(\mu\sqrt{t})
\left(\frac{\e^{-\mu^2t}}{\mu\sqrt{\pi t}}+\erf(\mu\sqrt{t})\right).
\label{pisca}
\eeqa
It is worthwhile recalling here the linear behaviour of the correlation
function,
\beq
C_n(t)\approx1-2\rho(t)\abs{n},
\eeq
in the so-called Porod regime ($\rho(t)\abs{n}\ll1$)~\cite{bray}.

The above expressions are functions of the scaling variable $\mu^2t\approx
t/\tau_\eq$.
The regimes where the latter variable is small or large deserve special
attention.

At zero temperature,
and more generally in the low-temperature coarsening regime
($\mu^2t\ll1$, i.e., $t\ll\tau_\eq$),
the correlation function assumes the scaling form
\beq
C_n(t)\approx\erfc\frac{\abs{n}}{2\sqrt{t}}.
\label{cgau}
\eeq
The growth law $L(t)\sim\sqrt{t}$ of the typical domain size
is another manifestation of the diffusive dynamics.
It is corroborated by the power-law behaviour of $\rho(t)$ and~$\chi(t)$:
\beq
\rho(t)\approx\frac{1}{2\sqrt{\pi t}},\qquad
\chi(t)\approx 4\sqrt\frac{t}{\pi},
\label{rczero}
\eeq
whose product converges to the non-trivial constant~\cite{gl2000}
\beq
\Pi_\coa=\frac{2}{\pi}.
\label{picoa}
\eeq
The non-exponential form~(\ref{cgau}) of the correlation function
and the non-triviality of the constant $\Pi_\coa$
are among the most salient non-equilibrium features of the zero-temperature
coarsening regime.

In the opposite regime ($\mu^2t\gg1$, i.e., $t\gg\tau_\eq$),
the correlation function $C_n(t)$ converges exponentially fast
to its equilibrium value~(\ref{ceqres}).
In particular, $\rho(t)$ and~$\chi(t)$ converge exponentially fast
to their equilibrium values~(\ref{rceq}), according~to
\beqa
\rho(t)&\approx&\frac{\mu}{2}\left(1+\frac{\e^{-\mu^2t}}{2\mu^3\sqrt{\pi
t^3}}+\cdots\right),
\label{rholar}
\\
\chi(t)&\approx&\frac{2}{\mu}\left(1-\frac{\e^{-\mu^2t}}{\mu\sqrt{\pi
t}}+\cdots\right),
\label{chilar}
\eeqa
so that their product converges to $\Pi_\eq=1$ (see~(\ref{pilar})).

\begin{figure}
\begin{center}
\includegraphics[angle=0,width=.7\linewidth,clip=true]{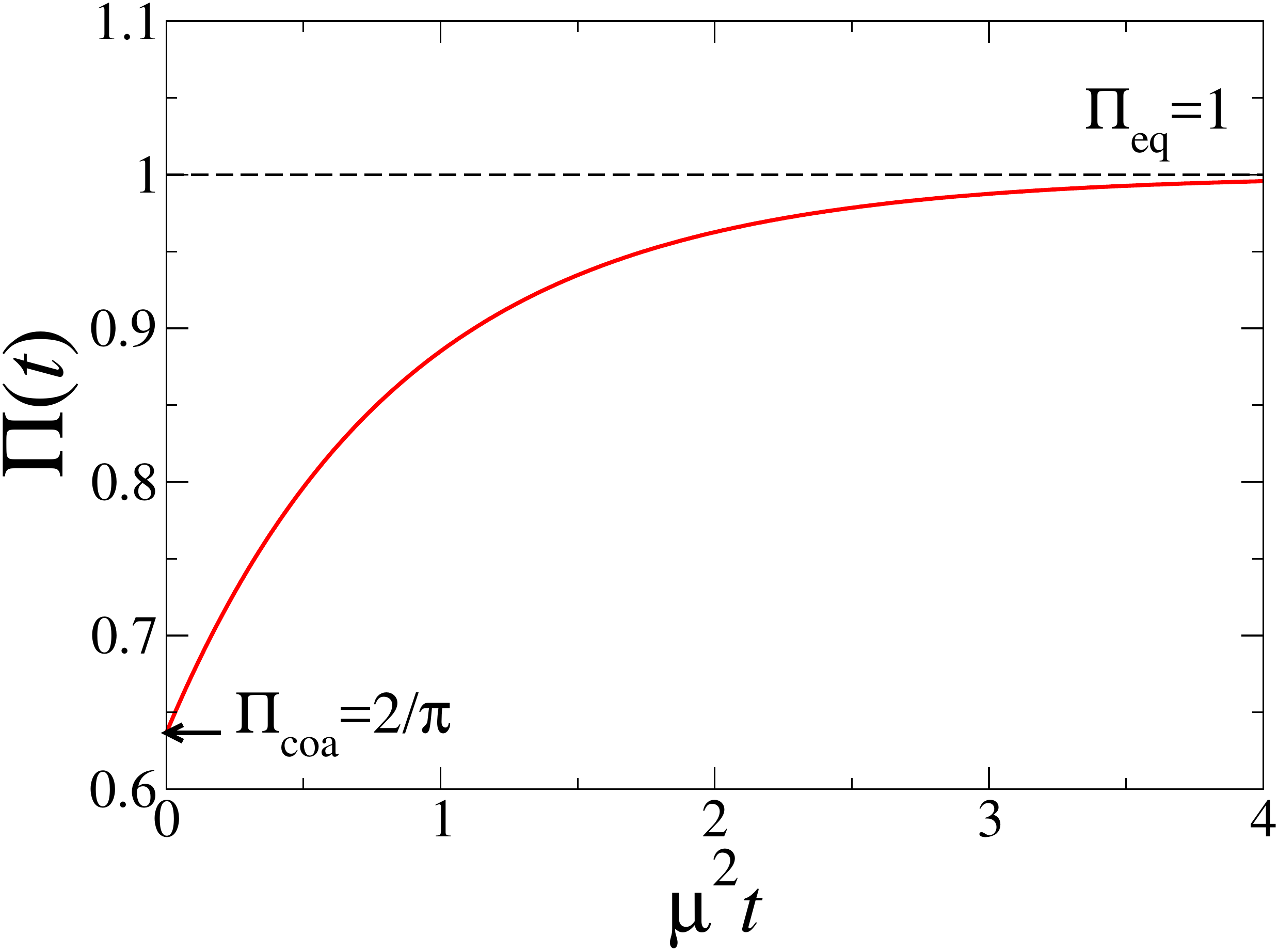}
\caption{\small
Form factor $\Pi(t)$ for a quench to a constant low temperature in the scaling regime
(see~(\ref{pisca})), against $\mu^2t$.}
\label{pilow}
\end{center}
\end{figure}

The form factor $\Pi(t)$ is plotted in figure~\ref{pilow} against $\mu^2t$.
It is an increasing function of the latter variable,
interpolating between the coarsening and equilibrium values $\Pi_\coa$ and
$\Pi_\eq$,
and behaving as
\beq
\Pi(t)\approx\frac{2}{\pi}\left(1+\frac{2\mu^2t}{3}+\cdots\right)
\qquad(\mu^2t\to0)
\eeq
and as
\beq
\Pi(t)\approx1-\frac{\e^{-\mu^2t}}{\mu\sqrt{\pi t}}
\left(1-\frac{1}{\mu^2t}+\cdots\right)\qquad(\mu^2t\to\infty).
\label{pilar}
\eeq

\subsection{The case of an arbitrary low-temperature protocol}
\label{lowgal}

In the case of an arbitrary protocol
in the low-temperature scaling regime,
specified by some given $\mu(t)$,
the solution to~(\ref{ccqdot}), with initial condition~(\ref{ccqinit}), reads
\beq
C^\F(q,t)=\e^{-q^2t-M(t)}
+\int_0^t\e^{-q^2(t-s)-(M(t)-M(s))}\,w(s)\dd s,
\label{ccsol}
\eeq
with
\beq
M(t)=\int_0^t\mu^2(s)\dd s.
\label{mdef}
\eeq
The condition~(\ref{ccqnor}) then yields
\beq
\frac{\e^{-M(t)}}{2\sqrt{\pi
t}}+\int_0^t\frac{\e^{-(M(t)-M(s))}}{2\sqrt{\pi(t-s)}}\,w(s)\dd s=1.
\label{cweq}
\eeq

The method to extract the source term $w(t)$ from the above integral equation
follows that used in section~\ref{fingal}.
We introduce the combination
\beq
\lambda(t)=\delta(t)+w(t),
\label{clamdef}
\eeq
so that~(\ref{cweq}) translates to
\beq
\int_0^t\frac{\e^{M(s)}}{2\sqrt{\pi(t-s)}}\,\lambda(s)\dd s=\e^{M(t)}.
\label{cwcon}
\eeq
In Laplace space,~(\ref{cwcon}) becomes
\beq\label{eq:1}
\frac{\ell_1(p)}{2\sqrt{p}}=\ell_0(p),
\eeq
where
\beqa
\ell_0(p)&=&\int_0^\infty\e^{M(t)-pt}\dd t,
\nonumber\\
\ell_1(p)&=&\int_0^\infty\e^{M(t)-pt}\lambda(t)\dd t,
\nonumber\\
\ell_2(p)&=&\int_0^\infty\e^{M(t)-pt}\mu^2(t)\dd t,
\label{l0l1l2}
\eeqa
hence~(\ref{mdef}) yields
\beq\label{eq:2}
\ell_2(p)=p\ell_0(p)-1.
\eeq
Eliminating $\ell_0(p)$ between~(\ref{eq:1}) and~(\ref{eq:2}), we are left with
\beq
\ell_1(p)=\frac{2}{\sqrt{p}}\;(\ell_2(p)+1).
\label{l1l2}
\eeq
The inverse Laplace transform of the first factor in the right-hand side reads
$2/\sqrt{\pi t}$.
We thus obtain the following expressions for the source term~$w(t)$
and the density of defects $\rho(t)$:
\beq
\rho(t)\approx\frac{w(t)}{4}
\approx\frac{\e^{-M(t)}}{2\sqrt\pi}
\left(\frac{1}{\sqrt{t}}+\int_0^t\frac{\e^{M(s)}}{\sqrt{t-s}}\,\mu^2(s)\dd
s\right),
\label{rhopro}
\eeq
where $M(t)$ is defined in~(\ref{mdef}).
The susceptibility is given by
(see~(\ref{ccsol}))
\beq
\chi(t)=C^\F(0,t)=\e^{-M(t)}+\int_0^t\e^{-(M(t)-M(s))}\,w(s)\dd s.
\eeq
The first term is negligible in the scaling regime.
Inserting the expression~(\ref{rhopro}) for the source term $w(t)$ into the
integral,
and changing the order of integration,
we are left with the formula
\beq
\chi(t)\approx\frac{4}{\sqrt\pi}\,\e^{-M(t)}
\left(\sqrt{t}+\int_0^t\sqrt{t-s}\,\e^{M(s)}\mu^2(s)\dd s\right).
\label{chipro}
\eeq

Equations~(\ref{rhopro}) and~(\ref{chipro}) are the key outcomes of this
section,
providing expressions for the two observables of central interest, $\rho(t)$
and $\chi(t)$,
for an arbitrary protocol in the low-temperature scaling regime.
These results are by far more tractable than their exact
finite-temperature analogues~(\ref{rhores}) and~(\ref{chires}).
In the case of a constant low temperature, where $\mu$ is a constant and
$M(t)=\mu^2t$,
the expressions~(\ref{rhosca}) and~(\ref{chisca}) are recovered.
In subsequent sections we shall employ~(\ref{rhopro}) and~(\ref{chipro})
to analyse diverse protocols in the low-temperature scaling regime.

\section{Everlasting slow quenches}
\label{ever}

The first protocol we shall investigate is that of an everlasting slow quench,
during which temperature decreases continuously, attaining zero only in the limit of
infinitely long times.
The scaling properties of the system
only depend on the tail of the quenching process,
namely how the parameter $\mu(t)$ goes to zero as $t\to\infty$.
We assume for definiteness that it falls off as a power law,
according to
\beq
\mu^2(t)\approx\frac{B}{t^\theta},
\label{mub}
\eeq
with arbitrary positive exponent $\theta$ and amplitude $B$.

A first picture of the scaling laws governing the system in the late-time
regime
can be obtained by means of the following heuristic reasoning,
in line with the Kibble-Zurek scaling theory~\cite{kibble,zurek}.
At thermal equilibrium, the relaxation time scales as $\tau_\eq\approx1/\mu^2$
(see~(\ref{taueq})).
In the present setting,
it is natural to introduce the instantaneous effective relaxation time
\beq
\tau_\eff(t)=\frac{1}{\mu^2(t)}.
\label{taudef}
\eeq
Equation~(\ref{mub}) implies the power-law growth $\tau_\eff(t)\sim t^\theta$.
So:

\begin{itemize}

\item
If $\theta>1$, the effective relaxation time is such that $\tau_\eff(t)\gg t$.
The system is therefore expected
to be asymptotically driven by an effective non-equilibrium zero-temperature
dynamics,
so that the scaling laws of the coarsening regime should apply for long times.

\item
If $\theta<1$, the effective relaxation time is such that $\tau_\eff(t)\ll t$.
The system is therefore expected
to follow adiabatically a finite-temperature dynamics
with a slowly varying time-dependent temperature,
and therefore to stay close to equilibrium at that temperature.

\item
If $\theta=1$, the system is in the marginal situation
where $t$ and $\tau_\eff(t)\approx t/B$ are proportional to each other.
This case deserves a more detailed analysis.

\end{itemize}

The qualitative predictions of the above line of thought turn out to be correct.
They will be turned into quantitative results in the remainder of this section.
It is to be noted that the resemblance of the above heuristic reasoning
with the Kibble-Zurek scaling theory is limited to the idea of comparing
relevant time scales.
The pointlike nature of elementary defects (domain walls) plays no role at all in this.

\subsection{Effective non-equilibrium zero-temperature dynamics}
\label{everfast}

This situation corresponds to everlasting slow quenches with $\theta>1$.
We have then $\mu(t)\ll1/\sqrt{t}$, and so $\tau_\eff(t)\gg t$.
This implies that the integral
\beq
M(\infty)=\int_0^\infty\mu^2(t)\dd t
\eeq
converges.
To leading order in the long-time regime,~(\ref{rhopro}) and~(\ref{chipro})
reduce to
\beqa
\rho(t)&\approx&\frac{\e^{-M(\infty)}}{2\sqrt{\pi t}}
\left(1+\int_0^\infty\e^{M(s)}\,\mu^2(s)\dd s\right),
\nonumber\\
\chi(t)&\approx&4\sqrt\frac{t}{\pi}\,\e^{-M(\infty)}
\left(1+\int_0^\infty\e^{M(s)}\,\mu^2(s)\dd s\right).
\eeqa
The integral equals $\e^{M(\infty)}-1$, so that we are left with
\beq
\rho(t)\approx\frac{1}{2\sqrt{\pi t}},\qquad
\chi(t)\approx 4\sqrt\frac{t}{\pi}.
\label{everrczero}
\eeq
The asymptotic power-law behaviour~(\ref{rczero}) of $\rho(t)$ and $\chi(t)$,
and the ensuing non-trivial limit $\Pi_\coa$ of their product
(see~(\ref{picoa})),
therefore hold unchanged for any everlasting quench such that $\theta>1$,
i.e., $\mu(t)\ll1/\sqrt{t}$.
These results corroborate the above picture in the style of Kibble-Zurek,
predicting that the system is driven by an effective non-equilibrium
zero-temperature dynamics.

The leading corrections to the asymptotic scaling laws~(\ref{everrczero})
can be obtained by rearranging and formally expanding~(\ref{rhopro})
and~(\ref{chipro}) as
\beqa
\rho(t)&\approx&\frac{1}{2\sqrt{\pi t}}
\left(1+\int_0^\infty\e^{M(s)-M(\infty)}
\left[\left(1-\frac{s}{t}\right)^{-1/2}-1\right]\,\mu^2(s)\dd s\right)
\nonumber\\
&\approx&\frac{1}{2\sqrt{\pi t}}
\left(1+\frac{m_1}{2t}+\cdots\right),
\nonumber\\
\chi(t)&\approx&4\sqrt\frac{t}{\pi}
\left(1+\int_0^\infty\e^{M(s)-M(\infty)}
\left[\left(1-\frac{s}{t}\right)^{1/2}-1\right]\,\mu^2(s)\dd s\right)
\nonumber\\
&\approx&4\sqrt\frac{t}{\pi}
\left(1-\frac{m_1}{2t}+\cdots\right),
\label{rcreg}
\eeqa
with the definition
\beq
m_k=\int_0^\infty\e^{M(s)-M(\infty)}\,s^k\,\mu^2(s)\dd s.
\eeq

If $\theta>2$, i.e., $\mu(t)\ll1/t$, the integral $m_1$ is convergent.
The leading corrections to~(\ref{everrczero}) are therefore
of relative order $1/t$, as given by~(\ref{rcreg}).
The situation is however different when $1<\theta<2$, i.e.,
$1/t\ll\mu(t)\ll1/\sqrt{t}$, where $m_1$ diverges.
We have then
\beqa
\rho(t)&\approx&\frac{1}{2\sqrt{\pi t}}
\left(1+\int_0^t
\left[\left(1-\frac{s}{t}\right)^{-1/2}-1\right]\,\mu^2(s)\dd s\right)
\nonumber\\
&\approx&\frac{1}{2\sqrt{\pi t}}
\left(1+\frac{K_1B}{t^{\theta-1}}+\cdots\right),
\nonumber\\
\chi(t)&\approx&4\sqrt\frac{t}{\pi}
\left(1+\int_0^t
\left[\left(1-\frac{s}{t}\right)^{1/2}-1\right]\,\mu^2(s)\dd s\right)
\nonumber\\
&\approx&4\sqrt\frac{t}{\pi}
\left(1-\frac{K_2B}{t^{\theta-1}}+\cdots\right),
\eeqa
with
\beqa
K_1
&=&\int_0^1\left(\frac{1}{\sqrt{1-u}}-1\right)u^{-\theta}\dd u
=\sqrt{\pi}\frac{\Gamma(1-\theta)}{\Gamma\left(\frac32-\theta\right)}+\frac{1}{\theta-1},
\nonumber\\
K_2
&=&\int_0^1\left(1-\sqrt{1-u}\right)u^{-\theta}\dd u
=-\sqrt{\pi}\frac{\Gamma(1-\theta)}{2\Gamma\left(\frac52-\theta\right)}-\frac{1}{\theta-1}.
\eeqa

The ensuing asymptotic expansion for $\Pi(t)$, namely
\beq
\Pi(t)=\frac{2}{\pi}\left(1+\frac{KB}{t^{\theta-1}}+\cdots\right),
\eeq
with
\beq
K=K_1-K_2=\sqrt{\pi}\frac{(2-\theta)\Gamma(1-\theta)}{\Gamma\left(\frac52-\theta\right)}
+\frac{2}{\theta-1},
\eeq
holds in the whole range $1<\theta<3$,
whereas for $\theta>3$ we have
\beq
\Pi(t)\approx\frac{2}{\pi}
\left(1+\frac{c}{4t^2}+\cdots\right),
\eeq
with $c=m_2-m_1^2$.

\subsection{Near-equilibrium finite-temperature dynamics}
\label{everslow}

This situation corresponds to everlasting slow quenches with $\theta<1$.
We have then $\mu(t)\gg1/\sqrt{t}$, and so $\tau_\eff(t)\ll t$.
The integral $M(t)$ grows as
\beq
M(t)\approx\frac{B}{1-\theta}\,t^{1-\theta}.
\eeq
As a consequence, the first terms in~(\ref{rhopro}) and~(\ref{chipro})
fall off exponentially in time.
Neglecting these contributions, we are left with the expressions
\beqa
\rho(t)&\approx&\frac{1}{2\sqrt{\pi}}
\int_0^t\frac{1}{\sqrt{t-s}}\,\e^{M(s)-M(t)}\,\mu^2(s)\dd s,
\nonumber\\
\chi(t)&\approx&\frac{4}{\sqrt{\pi}}
\int_0^t\sqrt{t-s}\;\e^{M(s)-M(t)}\,\mu^2(s)\dd s.
\label{rceff}
\eeqa
In the long-time regime,
the above integrals are dominated by times $s$ such that
\beq
\eps=t-s\ll t.
\eeq

To leading order, we have $\mu(t-\eps)\approx\mu(t)$
and $M(t-\eps)-M(t)\approx-\mu^2(t)\eps$,
and so~(\ref{rceff}) evaluate to
\beq
\rho(t)\approx\frac{\mu(t)}{2},\qquad\chi(t)\approx\frac{2}{\mu(t)}.
\label{rclead}
\eeq
The above formulas express the validity of the adiabatic approximation:
observables are nearly equal to their equilibrium values~(\ref{rceq})
at the instantaneous temperature~$\mu(t)$.
The ensuing form factor $\Pi(t)$ is accordingly very near its equilibrium value
$\Pi_\eq=1$ (see~(\ref{pie})).
These results, too, corroborate the above picture \`a la Kibble-Zurek,
predicting that the system is driven by an effective near-equilibrium
finite-temperature dynamics.

The outcome~(\ref{rclead}) of the adiabatic approximation
can be improved by means of the following systematic expansion.
Whenever $\mu(t)$ is slowly varying, we have
\beqa
\mu(t-\eps)&=&\mu(t)-\mu'(t)\eps+\frac{\mu''(t)}{2}\eps^2+\cdots,
\\
M(t-\eps)&=&M(t)
-\mu^2(t)\eps+\mu(t)\mu'(t)\eps^2-\frac{\mu(t)\mu''(t)+\mu'^2(t)}{3}\eps^3+\cdots
\nonumber
\eeqa
By inserting the above expansions into~(\ref{rceff}),
performing the integrals and rearranging terms,
we obtain the following expansions
around the adiabatic approximation for the observables of interest
in the case of an arbitrary slowly varying low temperature profile:
\beqa
\frac{2\rho(t)}{\mu(t)}&=&1-\frac{\mu'(t)}{4\mu^3(t)}
+\left(\frac{\mu''(t)}{8\mu^5(t)}-\frac{11\mu'^2(t)}{32\mu^6(t)}\right)+\cdots,
\nonumber\\
\frac{\chi(t)\mu(t)}{2}&=&1+\frac{3\mu'(t)}{4\mu^3(t)}
+\left(-\frac{5\mu''(t)}{8\mu^5(t)}+\frac{85\mu'^2(t)}{32\mu^6(t)}\right)+\cdots,
\nonumber\\
\Pi(t)&=&1+\frac{\mu'(t)}{2\mu^3(t)}
+\left(-\frac{\mu''(t)}{2\mu^5(t)}+\frac{17\mu'^2(t)}{8\mu^6(t)}\right)+\cdots
\label{fullslow}
\eeqa
The above expansions take slightly simpler forms if time derivatives of
$\mu(t)$
are expressed in terms of time derivatives of the effective relaxation time
$\tau_\eff(t)$
(see~(\ref{taudef})):
\beqa
\frac{2\rho(t)}{\mu(t)}&=&1+\frac{\tau_\eff'(t)}{8}
+\left(\frac{\tau_\eff'^2(t)}{128}-\frac{\tau_\eff(t)\tau_\eff''(t)}{16}\right)+\cdots,
\nonumber\\
\frac{\chi(t)\mu(t)}{2}&=&1-\frac{3\tau_\eff'(t)}{8}
+\left(\frac{25\tau_\eff'^2(t)}{128}+\frac{5\tau_\eff(t)\tau_\eff''(t)}{16}\right)+\cdots,
\nonumber\\
\Pi(t)&=&1-\frac{\tau_\eff'(t)}{4}
+\left(\frac{5\tau_\eff'^2(t)}{32}+\frac{\tau_\eff(t)\tau_\eff''(t)}{4}\right)+\cdots
\label{fulltauslow}
\eeqa
In the case where $\mu(t)$ is (exactly) given by the power law~(\ref{mub}) with $\theta<1$,
the above expansions become asymptotic expansions in powers of $t^{-(1-\theta)}$:
\beqa
\frac{2\rho(t)}{\mu(t)}&=&
1+\frac{\theta}{8B}\,t^{-(1-\theta)}
+\frac{\theta(8-7\theta)}{128B^2}\,t^{-2(1-\theta)}+\cdots,
\nonumber\\
\frac{\chi(t)\mu(t)}{2}&=&
1-\frac{3\theta}{8B}\,t^{-(1-\theta)}
+\frac{5\theta(13\theta-8)}{128B^2}\,t^{-2(1-\theta)}+\cdots,
\nonumber\\
\Pi(t)&=&1-\frac{\theta}{4B}\,t^{-(1-\theta)}
+\frac{\theta(13\theta-8)}{32B^2}\,t^{-2(1-\theta)}+\cdots
\eeqa

\subsection{The marginal case}
\label{evermar}

We close this study of everlasting quenches by considering the marginal case
where the parameter $\mu(t)$ falls off as
\beq
\mu^2(t)\approx\frac{B}{t},
\label{mumar}
\eeq
with an arbitrary amplitude $B$.
The integral $M(t)$ therefore grows as
\beq
M(t)\approx B\,\ln\frac{t}{t_0},
\label{mmar}
\eeq
where the microscopic time scale $t_0$ depends on details of the behaviour
of $\mu(t)$ at finite times.
As a consequence, the first terms in~(\ref{rhopro}) and~(\ref{chipro})
decay much faster than the second ones,
as they are reduced by a factor $\e^{-M(t)}\approx(t_0/t)^B$.
Neglecting these contributions, we are again left with the
expressions~(\ref{rceff}).

Inserting~(\ref{mumar}) and~(\ref{mmar}) into~(\ref{rceff}) and performing the integrals,
we obtain the scaling laws
\beq
\rho(t)\approx\frac{X(B)}{2\sqrt{\pi t}},\qquad
\chi(t)\approx 4\sqrt\frac{t}{\pi}\,Y(B).
\label{rcmar}
\eeq
The power laws~(\ref{rczero}) of $\rho(t)$ and $\chi(t)$,
characteristic of zero-temperature coarsening dynamics,
are affected by multiplicative factors $X(B)$ and $Y(B)$,
which depend continuously on the amplitude $B$, according to
\beq
X(B)=\sqrt{\pi}\,\frac{\Gamma(B+1)}{\Gamma(B+\frac12)},\qquad
Y(B)=\sqrt{\pi}\,\frac{\Gamma(B+1)}{2\Gamma(B+\frac32)}.
\eeq
The form factor $\Pi(t)$ therefore converges to the following limit:
\beq
\Pi(B)=\frac{2}{\pi}\,X(B)Y(B)=\frac{\Gamma^2(B+1)}{\Gamma(B+\frac12)\Gamma(B+\frac32)},
\label{pib}
\eeq
which also depends continuously on $B$, increasing monotonically
from $\Pi_\coa=2/\pi$ at $B=0$ to $\Pi_\eq=1$ as $B\to\infty$.
When $B$ is an integer, $\Pi(B)$ admits the product formula
\beq
\Pi(B)=\frac{2}{\pi}\prod_{k=1}^B\frac{4k^2}{(2k-1)(2k+1)},
\eeq
which has attracted some interest lately, in connection with the Wallis
formula~\cite{wallis}.

When the amplitude $B$ is small,
the above results exhibit smooth corrections
with respect to the non-equilibrium zero-temperature coarsening regime:
\beqa
X(B)&=&1+2\ln 2\,B+\cdots,
\nonumber\\
Y(B)&=&1+2(\ln 2-1)B+\cdots,
\nonumber\\
\Pi(B)&=&\frac{2}{\pi}\Bigl(1+2(2\ln 2-1)B+\cdots\Bigr).
\eeqa

When the amplitude $B$ is large,
the above results manifest a crossover towards
the equilibrium finite-temperature regime.
We have
\beqa
\frac{2\rho(t)}{\mu(t)}
\approx\frac{X(B)}{\sqrt{\pi B}}
=\frac{\Gamma(B+1)}{\sqrt{B}\,\Gamma(B+\frac12)}
=1+\frac{1}{8B}+\frac{1}{128B^2}\cdots,
\\
\frac{\chi(t)\mu(t)}{2}
\approx2\sqrt\frac{B}{\pi}\,Y(B)
=\frac{\sqrt{B}\,\Gamma(B+1)}{\Gamma(B+\frac32)}
=1-\frac{3}{8B}+\frac{25}{128B^2}+\cdots
\nonumber
\eeqa
and
\beq
\Pi(B)=1-\frac{1}{4B}+\frac{5}{32B^2}+\cdots
\eeq
A much longer asymptotic expansion of $\Pi(B)$ is given
in~\cite[Eq.~(3.12)]{wallis}.

\section{Slow quenches of finite duration}
\label{slofin}

\begin{figure}
\begin{center}
\includegraphics[angle=0,width=.6\linewidth,clip=true]{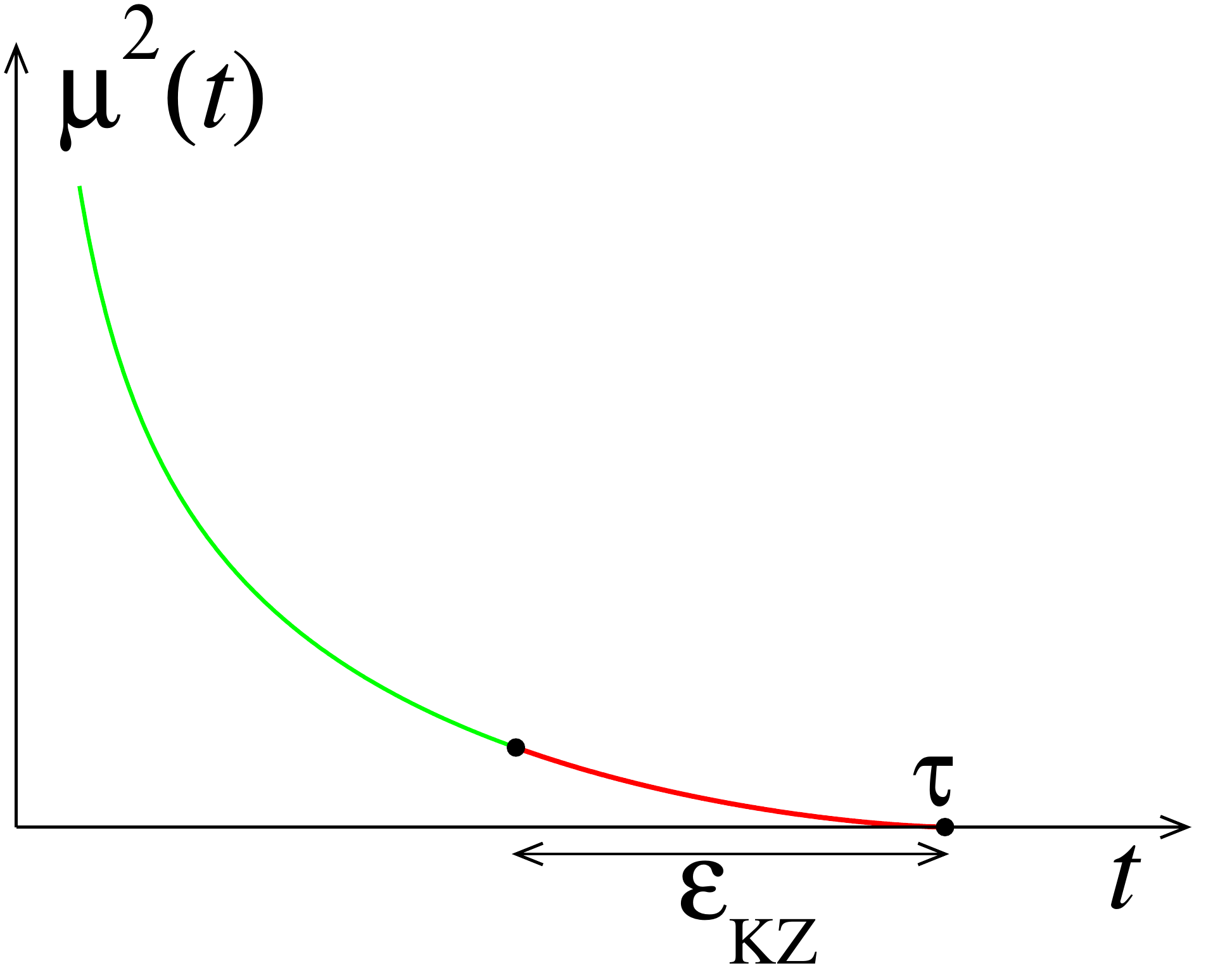}
\caption{\small
Sketchy plot of the time dependence of $\mu^2(t)$ for a slow quench of finite duration.
The Kibble-Zurek time scale $\eps_\kz$ is introduced in~(\ref{kzdef}).}
\label{qdeux}
\end{center}
\end{figure}

The next protocol we consider is that of a slow quench over a finite time interval,
during which temperature decreases continuously,
reaches $T(\tau)=0$ at some long but finite quenching time $\tau$,
and then stays equal to zero, as sketched in figure~\ref{qdeux}.

This protocol has attracted much
attention in a recent past~\cite{brey,krapiv,jeong,priyanka,mayo}.
These earlier works are focussed on the behaviour of density of defects $\rho(\tau)$
right at the quenching time $\tau$,
in the regime of most interest where the latter time is large.
In this regime, the scaling properties of the system
only depend on the tail of the slow quenching process,
namely how $T(t)$ goes to zero as $t\to\tau$.
We choose to parametrize the latter tail as
\beq
\mu^2(\tau-\eps)\approx
A\left(\frac{\eps}{\tau}\right)^\alpha\qquad(\eps\ll\tau),
\label{slomu}
\eeq
with arbitrary positive exponent $\alpha$ and amplitude $A$.
Henceforth we shall successively consider the system at the quenching time
($t=\tau$)
and in the late-time regime ($t\gg\tau$).

\subsection{Scaling properties at the quenching time}
\label{sloque}

The purpose of this section is to determine the scaling behaviour
of the density of defects $\rho(\tau)$
and of the susceptibility $\chi(\tau)$ right at the quenching time $\tau$,
in the regime where $\tau$ is large.
This can be done by inserting the expression~(\ref{slomu}) of $\mu^2(t)$,
as well as the ensuing estimate
\beq
M(\tau)-M(\tau-\eps)\approx\frac{A\,\eps^{\alpha+1}}{(\alpha+1)\tau^\alpha}
\qquad(\eps\ll\tau),
\label{slom}
\eeq
into the expressions~(\ref{rhopro}) and~(\ref{chipro}).
The latter formulas thus simplify to
\beq
\rho(\tau)\approx\frac{1}{2\sqrt\pi}
\int_0^\infty\frac{f(\eps)}{\sqrt{\eps}}\,\dd\eps,\quad
\chi(\tau)\approx\frac{4}{\sqrt\pi}
\int_0^\infty\sqrt\eps\,f(\eps)\,\dd\eps,
\eeq
with
\beq
f(\eps)=A\left(\frac{\eps}{\tau}\right)^\alpha
\exp\left(-\frac{A\,\eps^{\alpha+1}}{(\alpha+1)\tau^\alpha}\right).
\label{fden}
\eeq
We thus obtain the explicit results
\beqa
\rho(\tau)
&\approx&\frac{1}{2\sqrt\pi}\;\Gamma\!\left(\frac{2\alpha+1}{2\alpha+2}\right)
\left(\frac{A}{(\alpha+1)\tau^\alpha}\right)^{1/(2(\alpha+1))},
\label{rslo}
\\
\chi(\tau)
&\approx&\frac{4}{\sqrt\pi}\;\Gamma\!\left(\frac{2\alpha+3}{2\alpha+2}\right)
\left(\frac{(\alpha+1)\tau^\alpha}{A}\right)^{1/(2(\alpha+1))}.
\label{cslo}
\eeqa

The exponent
\beq
\lambda=\frac{\alpha}{2(\alpha+1)},
\eeq
characterizing the power-law behaviour of the typical domain size at the
quenching
time~$\tau$,
\beq
\frac{1}{\rho(\tau)}\sim\chi(\tau)\sim\tau^\lambda,
\label{scalam}
\eeq
is an increasing function of $\alpha$ in the range $0<\lambda<1/2$.
Its expression was obtained in~\cite{brey,krapiv,jeong,priyanka},
and shown to agree with the predictions of the Kibble-Zurek scaling
theory~\cite{kibble,zurek}.
The reasoning goes as follows.
The effective relaxation time introduced in~(\ref{taudef})
continuously increases as the end of the quench is approached.
It becomes equal to the remaining time $\eps$ until the end of the quench
when $\eps$ crosses the Kibble-Zurek time scale
\beq
\eps_\kz\approx\left(\frac{\tau^\alpha}{A}\right)^{1/(\alpha+1)},
\label{kzdef}
\eeq
such that $\mu^2(\eps_\kz)\approx1/\eps_\kz$.
It is therefore expected that the system first remains close to thermal
equilibrium
(green part of the plot in figure~\ref{qdeux}),
until it falls out of equilibrium when $\tau_\eff(t)\approx\tau-t$,
i.e., $t\approx\tau-\eps_\kz$,
and remains out of equilibrium until the end of the quench
(red part of the plot).
The Kibble-Zurek scaling theory therefore predicts
\beq
1/\rho(\tau)\sim\chi(\tau)\sim1/\mu(\eps_\kz)\sim\sqrt{\eps_\kz}\sim\tau^\lambda,
\eeq
in agreement with~(\ref{scalam}).

The above quantitative analysis is by far simpler than those
of~\cite{brey,krapiv,jeong,priyanka}.
The result~(\ref{cslo}) for~$\chi(t)$ is of course novel,
whereas the exact amplitude of $\rho(\tau)$ in~(\ref{rslo}) was first derived
in~\cite{priyanka},
starting from the finite-temperature formalism of~\cite{brey}.
Surprisingly enough,
an earlier erroneous prediction of the amplitude of $\rho(\tau)$~\cite{krapiv}
turns out to coincide with the exact amplitude~(\ref{cslo}) for $1/\chi(\tau)$.

As a consequence of~(\ref{rslo}),~(\ref{cslo}),
the form factor $\Pi(\tau)$ only depends on the quenching exponent $\alpha$,
just as the growth exponent $\lambda$, according to
\beq
\Pi(\tau)=\frac{1}{(\alpha+1)\sin\frad{\pi}{2(\alpha+1)}}=\frac{1-2\lambda}{\cos\pi\lambda}.
\label{pislo}
\eeq
It is a decreasing function of $\alpha$,
starting from its equilibrium value $\Pi_\eq=1$ in the limit of a very steep
slow quench
($\alpha\to0$, $\lambda\to0$):
\beq
\Pi(\tau)=1-\alpha+\left(\frac{\pi^2}{8}+1\right)\alpha^2+\cdots,
\eeq
and going towards its coarsening value $\Pi_\coa=2/\pi$ in the limit of a very
mild slow quench
($\alpha\to\infty$, $\lambda\to1/2$):
\beq
\Pi(\tau)=\frac{2}{\pi}+\frac{\pi}{12\alpha^2}-\frac{\pi}{6\alpha^3}+\cdots
\eeq
In the special case of a linear temperature ramp,
considered in detail in~\cite{krapiv,jeong,priyanka},
and more generally of a quenching profile ending linearly ($\alpha=1$),
we have
\beq
\lambda=\frac{1}{4},\qquad\Pi(\tau)=\frac{1}{\sqrt{2}}.
\eeq

\subsection{Scaling properties in the long-time regime}
\label{slolon}

The purpose of this section is to analyse the scaling behaviour
of the density of defects~$\rho(t)$
and of the susceptibility $\chi(t)$
in the long-time regime ($t\gg\tau$),
long after the end of a slow quench of finite duration.

To leading order in the long-time regime,~(\ref{rhopro}) and~(\ref{chipro})
reduce to
\beqa
\rho(t)&\approx&\frac{\e^{-M(\tau)}}{2\sqrt{\pi t}}
\left(1+\int_0^\tau\e^{M(s)}\,\mu^2(s)\dd s\right),
\nonumber\\
\chi(t)&\approx&4\sqrt\frac{t}{\pi}\,\e^{-M(\tau)}
\left(1+\int_0^\tau\e^{M(s)}\,\mu^2(s)\dd s\right).
\label{rclong}
\eeqa
We have indeed $M(t)=M(\tau)$ for all $t\ge\tau$.
The subsequent analysis follows that of section~\ref{everfast}.
The integral entering~(\ref{rclong}) equals $\e^{M(\tau)}-1$, so that we are
left with
\beq
\rho(t)\approx\frac{1}{2\sqrt{\pi t}},\qquad
\chi(t)\approx 4\sqrt\frac{t}{\pi}.
\eeq
The asymptotic power-law behaviours~(\ref{rczero}) of $\rho(t)$ and $\chi(t)$,
and the ensuing non-trivial limit $\Pi_\coa$ of their product
(see~(\ref{picoa})),
which were obtained for zero-temperature dynamics, i.e., an instantaneous
quench,
hold unchanged for any progressive quench of finite duration.

The corrections to the above leading asymptotic behaviour
can be obtained by rearranging and expanding~(\ref{rhopro}) and~(\ref{chipro}) as
\beqa
\rho(t)&\approx&\frac{1}{2\sqrt{\pi t}}
\left(1+\int_0^\tau\e^{M(s)-M(\tau)}
\left[\left(1-\frac{s}{t}\right)^{-1/2}-1\right]\,\mu^2(s)\dd s\right)
\nonumber\\
&\approx&\frac{1}{2\sqrt{\pi t}}
\left(1+\frac{m_1}{2t}+\frac{3m_2}{8t^2}+\cdots\right),
\nonumber\\
\chi(t)&\approx&4\sqrt\frac{t}{\pi}
\left(1+\int_0^\tau\e^{M(s)-M(\tau)}
\left[\left(1-\frac{s}{t}\right)^{1/2}-1\right]\,\mu^2(s)\dd s\right)
\nonumber\\
&\approx&4\sqrt\frac{t}{\pi}
\left(1-\frac{m_1}{2t}-\frac{m_2}{8t^2}+\cdots\right),
\label{rcasy}
\eeqa
with the definition
\beq
m_k=\int_0^\tau\e^{M(s)-M(\tau)}\,s^k\,\mu^2(s)\dd s.
\label{momsdef}
\eeq
We have therefore
\beq
\Pi(t)\approx\frac{2}{\pi}
\left(1+\frac{c}{4t^2}+\cdots\right),
\label{piasy}
\eeq
with $c=m_2-m_1^2$.
The asymptotic expansions~(\ref{rcasy}),~(\ref{piasy})
hold for any progressive quench of finite duration.

For a slow quench parametrized as~(\ref{slomu}) and a long quenching time $\tau$,
the leading-order growth of the moments $m_k$ reads
\beq
m_k\approx\tau^k.
\eeq
The first corrections in~(\ref{rcasy}) therefore simplify to
\beqa
\rho(t)&\approx&\frac{1}{2\sqrt{\pi t}}\left(1+\frac{\tau}{2t}+\cdots\right)
\approx\frac{1}{2\sqrt{\pi(t-\tau)}},
\nonumber\\
\chi(t)&\approx&4\sqrt\frac{t}{\pi}\left(1-\frac{\tau}{2t}+\cdots\right)
\approx4\sqrt\frac{t-\tau}{\pi}.
\eeqa
The above expressions show that the long-time behaviour of the system
is as if it had been subjected to an instantaneous quench at time $\tau$,
irrespective of the quenching exponent $\alpha$.

The evaluation of the second-order amplitude $c$ entering~(\ref{piasy}) is more
delicate.
A careful expansion of the integral entering the definition~(\ref{momsdef})
yields
\beq
c=m_2-m_1^2=\mean{\eps^2}_f-\mean{\eps}_f^2,
\eeq
with the definition
\beq
\mean{\eps^k}_f=\int_0^\infty\eps^k\,f(\eps)\,\dd\eps
\eeq
(see~(\ref{fden})),
and so
\beq
c=\left(\Gamma\!\left(\frac{\alpha+3}{\alpha+1}\right)
-\Gamma\!\left(\frac{\alpha+2}{\alpha+1}\right)^2\right)
\left(\frac{(\alpha+1)\tau^\alpha}{A}\right)^{2/(\alpha+1)}.
\eeq
This amplitude grows as $c\sim\tau^{4\lambda}\sim\eps_\kz^2$.

\section{Time-periodic protocols}
\label{per}

Besides the slow quenches considered so far,
many other classes of low-temperature heating and cooling protocols
can be studied by means of the scaling analysis of section~\ref{low}.
In this section we investigate protocols where the temperature parameter~$\mu(t)$
is modulated periodically in time.
To our knowledge, the consideration of such periodic temperature protocols
for the Glauber-Ising chain is novel.
The zero-temperature Glauber-Ising chain under a homogeneous time-periodic magnetic field
has however been investigated recently~\cite{YB}.

We consider for definiteness the harmonic protocol defined by
\beq
\mu^2(t)=a+b\cos\omega t,
\label{muper}
\eeq
with $a\ge b\ge0$.
The three positive parameters $a$, $b$ and $\omega$ have the dimension of a
frequency,
in our system of reduced units.
They are assumed to be small, in order to be in the low-temperature scaling
regime.
The integral $M(t)$ reads (see~(\ref{mdef}))
\beq
M(t)=at+\frac{b}{\omega}\,\sin\omega t.
\eeq

\subsection{General analysis}

Hereafter we shall not consider the transient regime,
and focus our attention on the stationary regime,
where observables are locked with the driving temperature protocol,
and therefore vary periodically in time with the same period.
In this regime, setting $s=t-\eps$,
with a positive time lag $\eps$,
the formulas~(\ref{rhopro}) and~(\ref{chipro}) become
\beqa
\rho(t)&\approx&\int_0^\infty\frac{1}{2\sqrt{\pi\eps}}\,\e^{M(t-\eps)-M(t)}\,\mu^2(t-\eps)\dd\eps,
\nonumber\\
\chi(t)&\approx&\int_0^\infty4\sqrt\frac{\eps}{\pi}\,\e^{M(t-\eps)-M(t)}\,\mu^2(t-\eps)\dd\eps.
\label{rcper}
\eeqa

Within this setting, it is suitable to represent periodic functions as Fourier
series.
We have
\beq
\e^{M(t)}=\e^{at}\e^{(b/\omega)\sin\omega t}=\e^{at}\sum_kE_k\,\e^{\ii k\omega
t},
\eeq
with
\beq
E_k=\int_0^{2\pi}\frac{\dd\theta}{2\pi}\,\e^{-\ii
k\theta+(b/\omega)\sin\theta}=\ii^{-k}\,I_k(b/\omega),
\eeq
where the $I_k$ are modified Bessel functions (see~\ref{appb}),
and therefore
\beq
\e^{M(t-\eps)-M(t)}=\e^{-a\eps}\sum_{k,l}E_kE_l\,\e^{\ii(k-l)\omega
t}\,\e^{-\ii k\omega\eps}.
\label{dble}
\eeq

Inserting the expression~(\ref{muper})
and the double series expansion~(\ref{dble}) into~(\ref{rcper}),
performing the integrals,
and using the identity (see~(\ref{bdiff}))
\beq
aE_k+\frac{b}{2}(E_{k-1}+E_{k+1})=(a+\ii k\omega)E_k,
\eeq
we obtain the Fourier series of $\rho(t)$ and $\chi(t)$ in the form
\beqa
\rho(t)&=&\sum_k\rho_k\,\e^{\ii k\omega t},\qquad
\rho_k=\sum_m\frac{\sqrt{a+\ii m\omega}}{2}\,E_mE_{m-k},
\nonumber\\
\chi(t)&=&\sum_k\chi_k\,\e^{\ii k\omega t},\qquad
\chi_k=\sum_m\frac{2}{\sqrt{a+\ii m\omega}}\,E_mE_{m-k}.
\label{rcser}
\eeqa

\subsection{Linear-response regime}

In the linear-response regime where the modulation amplitude $b$ is small,
all observables are expected to exhibit small harmonic oscillations around
their mean values, with amplitudes proportional to $b$.

In the case of the observables under study, $\rho(t)$, $\chi(t)$ and $\Pi(t)$,
this regime can be investigated analytically
from the general formulas~(\ref{rcser}).
To first order in $b$,
the non-zero Fourier coefficients $E_k$, $\rho_k$ and $\chi_k$ read
\beqa
E_0&=&1,\qquad
E_{\pm1}=\mp\frac{\ii b}{2\omega},
\nonumber\\
\rho_0&=&\frac{\mu_0}{2},\qquad
\rho_{\pm1}=\pm\frac{\ii b}{4\omega}\left(\mu_0-\sqrt{a\pm\ii\omega}\right),
\nonumber\\
\chi_0&=&\frac{2}{\mu_0},\qquad
\chi_{\pm1}=\pm\frac{\ii
b}{\omega}\left(\frac{1}{\mu_0}-\frac{1}{\sqrt{a\pm\ii\omega}}\right),
\eeqa
with the notation
\beq
\mu_0=\sqrt{a}.
\eeq
Introducing the angle $\phi$ such that
\beq
\tan\phi=\frac{\omega}{a}\qquad(0<\phi<\pi/2),
\label{phidef}
\eeq
we have
\beq
\sqrt{a\pm\ii\omega}=\mu_0\frac{\e^{\pm\ii\phi/2}}{\sqrt{\cos\phi}}.
\eeq

We thus obtain the following expressions in the linear-response regime,
for arbitrary frequencies:
\beqa
\rho(t)\approx\frac{\mu_0}{2}\left(1+\frac{b}{a}F_\rho(\omega t)\right),
\nonumber\\
\chi(t)\approx\frac{2}{\mu_0}\left(1+\frac{b}{a}F_\chi(\omega t)\right),
\nonumber\\
\Pi(t)\approx1+\frac{b}{a}F_\Pi(\omega t),
\eeqa
where the response functions read
\beqa
F_\rho(\omega t)=\frac{1}{\tan\phi}
\left(\frac{\sin(\omega t+\phi/2)}{\sqrt{\cos\phi}}-\sin\omega t\right),
\nonumber\\
F_\chi(\omega t)=-\sqrt{\cos\phi}\,F_\rho(\omega t-\phi/2),
\nonumber\\
F_\Pi(\omega t)=F_\rho(\omega t)+F_\chi(\omega t).
\label{fper}
\eeqa
The three response functions are semi-periodic functions,
obeying $F(\omega t+\pi)=-F(\omega t)$.
Their amplitudes, i.e., their maximal values over one period,
are respectively given by
\beqa
A_\rho=\frac{\sqrt{\cos\phi}}{\sin\phi}
\left(1+\cos\phi-2\sqrt{\cos\phi}\cos(\phi/2)\right)^{1/2},
\nonumber\\
A_\chi=\sqrt{\cos\phi}\,A_\rho,
\nonumber\\
A_\Pi=\frac{2\cos(\phi/2)\sqrt{\cos\phi}}{\sin\phi}
\left(\cos(\phi/2)-\sqrt{\cos\phi}\right).
\label{aper}
\eeqa

At low frequency $(\omega\ll a,\phi\to0)$, the results~(\ref{fper}) can be
expanded as
\beqa
F_\rho(\omega t)=\frac{\cos\omega t}{2}+\frac{\omega\,\sin\omega
t}{8a}-\frac{\omega^2\cos\omega t}{16a^2}+\cdots,
\nonumber\\
F_\chi(\omega t)=-\frac{\cos\omega t}{2}-\frac{3\o\,\sin\omega
t}{8a}+\frac{5\omega^2\cos\omega t}{16a^2}+\cdots,
\nonumber\\
F_\Pi(\omega t)=-\frac{\omega\,\sin\omega t}{4a}+\frac{\omega^2\cos\omega
t}{4a^2}+\cdots
\eeqa
These expansions are in full agreement with~(\ref{fullslow}).
As far as $\rho(t)$ and~$\chi(t)$ are concerned,
the first terms agree with the adiabatic approximation~(\ref{rclead}).
The second terms, proportional to frequency $\omega$,
manifest a retardation effect of the form
\beq
\rho(t)\approx\frac{\mu(t-\tau_\rho)}{2},\qquad
\chi(t)\approx\frac{2}{\mu(t-\tau_\chi)},
\eeq
where the retardation times $\tau_\rho$ and $\tau_\chi$ read
\beq
\tau_\rho\approx\frac{1}{4a}\approx\frac{\tau_0}{4},\qquad
\tau_\chi\approx\frac{3}{4a}\approx\frac{3\tau_0}{4},
\eeq
where $\tau_0\approx1/a$ is the equilibrium relaxation time in the absence of modulation
(see~(\ref{taueq})).
The retardation time $\tau_\chi$ is three times larger than~$\tau_\rho$.

At high frequency $(\omega\gg a,\phi\to\pi/2)$, we have
\beqa
F_\rho(t)\approx\sqrt\frac{a}{\omega}\cos(\omega t-\pi/4),
\nonumber\\
F_\chi(t)\approx-\frac{a}{\omega}\sin(\omega t).
\label{fhi}
\eeqa
The response functions fall off as powers of frequency, with different
exponents.
The decay of the non-local observable $\chi(t)$
is much faster than that of the local observable~$\rho(t)$.

\begin{figure}
\begin{center}
\includegraphics[angle=0,width=.7\linewidth,clip=true]{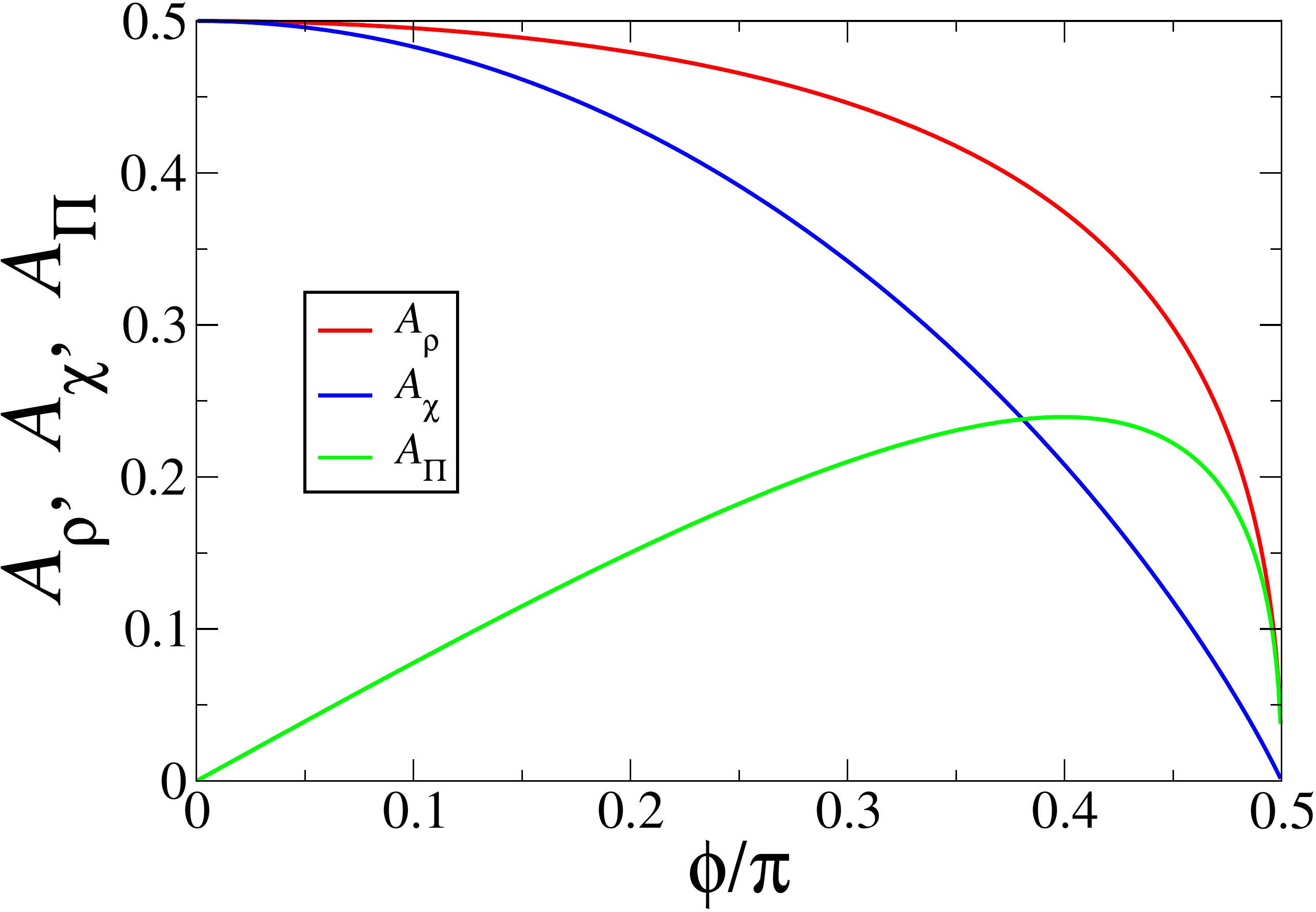}
\caption{\small
Amplitudes $A_\rho$, $A_\chi$ and $A_\Pi$ (see~(\ref{aper}))
of the response functions of the periodic protocol~(\ref{muper})
in the linear-response regime,
against $\phi/\pi$~(see~(\ref{phidef})).}
\label{amps}
\end{center}
\end{figure}

The above results are illustrated in figure~\ref{amps},
showing the amplitudes $A_\rho$, $A_\chi$ and $A_\Pi$ (see~(\ref{aper}))
against $\phi/\pi$.
The amplitudes $A_\rho$ and $A_\chi$ are monotonically decreasing
from their adiabatic values $1/2$ as $\phi\to0$ (i.e., $\omega\ll a$)
to zero as $\phi\to\pi/2$ (i.e., $\omega\gg a$).
In agreement with~(\ref{fhi}), $A_\chi$ vanishes linearly,
whereas $A_\rho$ ends up with a square-root singularity.
The amplitude $A_\Pi$ of the form factor vanishes both at low and at high frequency,
and reaches a non-trivial maximal value $A_\pi\approx0.239392$ at
$\phi\approx1.1254437$, i.e., $\omega/a\approx3.054807$.

\subsection{Critical modulation}

We now turn to the case where the modulation is the strongest, namely $b=a$.
This situation is critical, in the sense that the temperature parameter
\beq
\mu^2(t)=a(1+\cos\omega t)=2a\cos^2\frac{\omega t}{2}
\label{strong}
\eeq
vanishes in the middle of each period.

\begin{figure}
\begin{center}
\includegraphics[angle=0,width=.7\linewidth,clip=true]{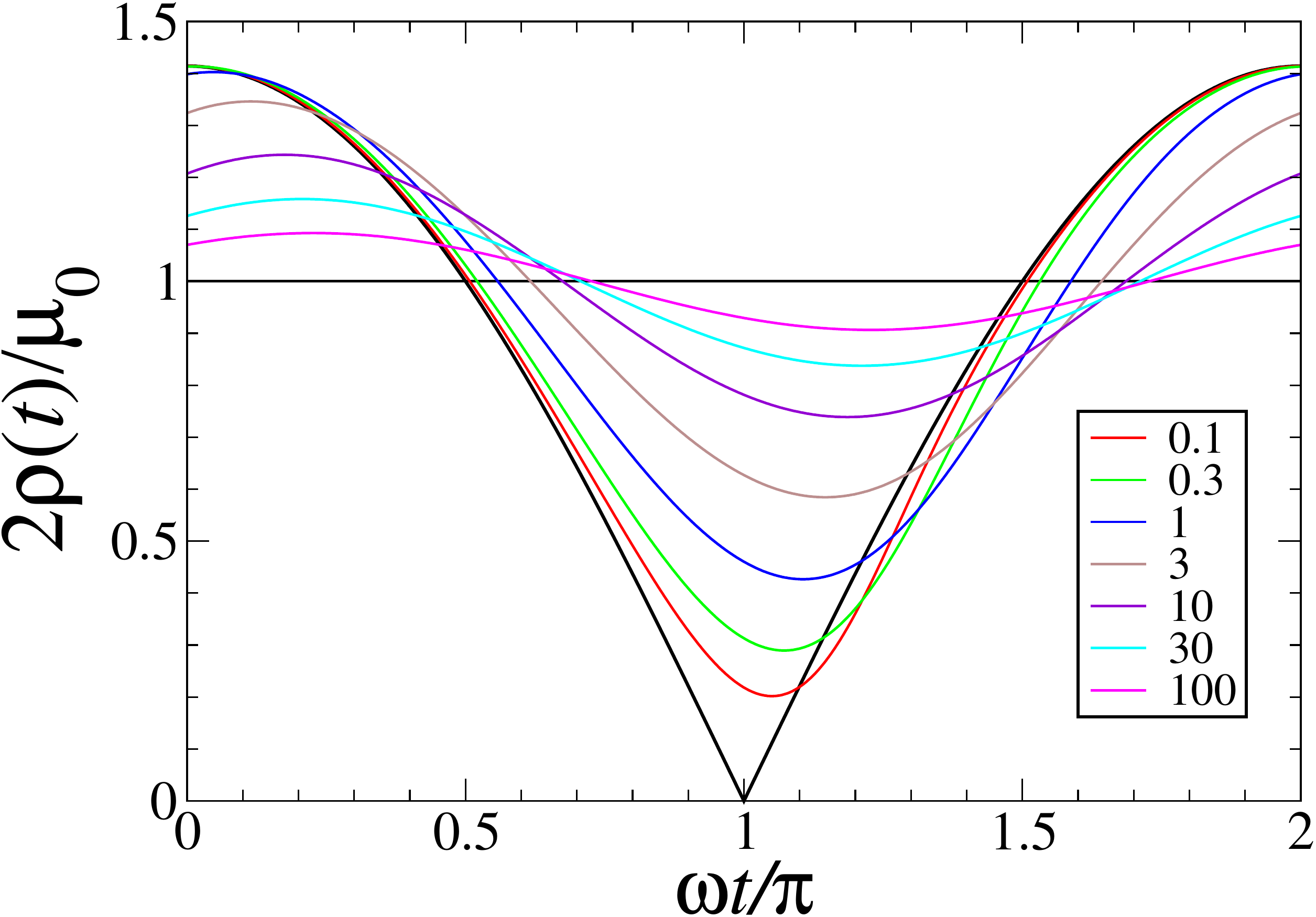}

\includegraphics[angle=0,width=.7\linewidth,clip=true]{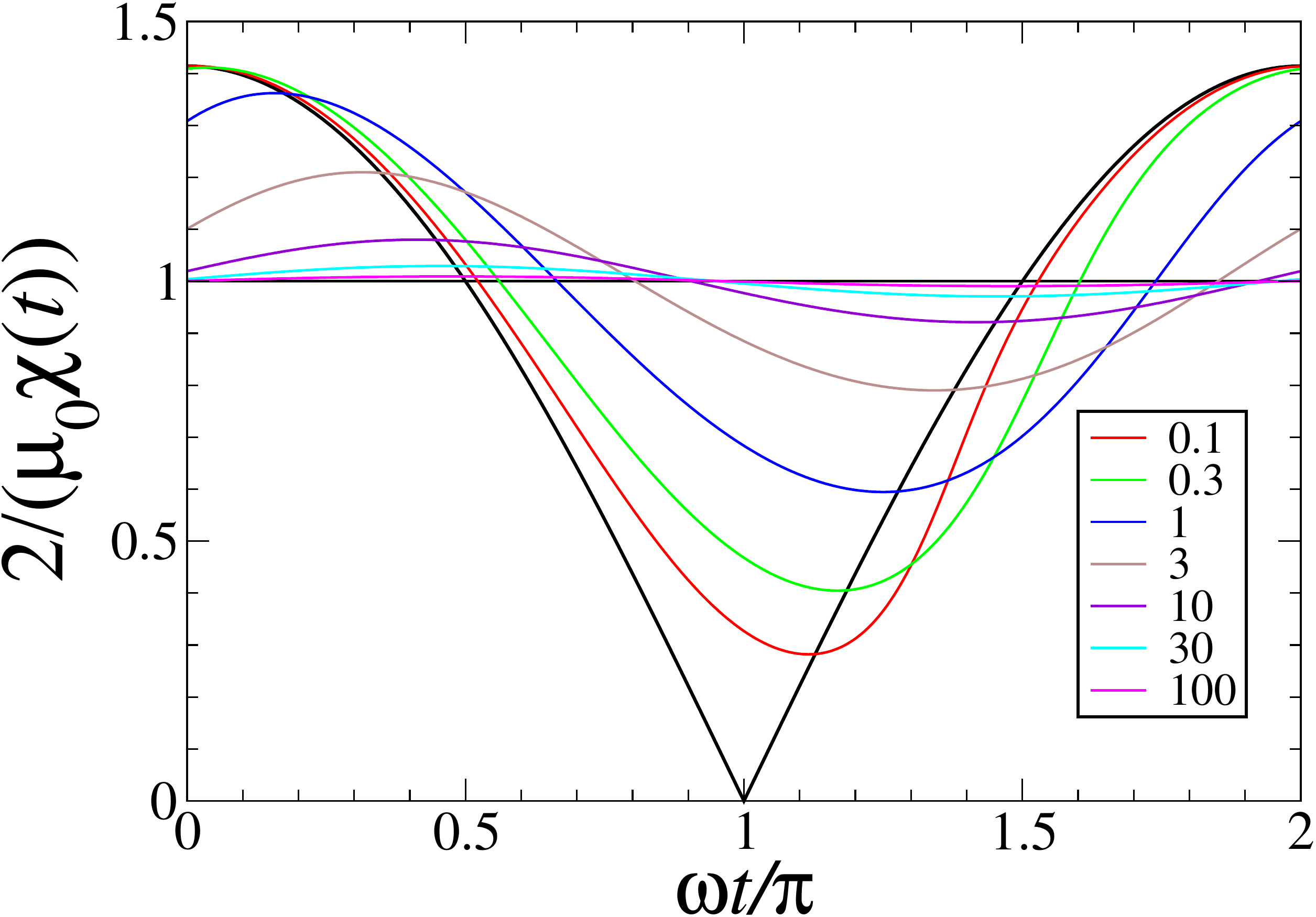}
\caption{\small
Reduced quantities $2\rho(t)/\mu_0$ (top) and $2/(\mu_0\chi(t))$ (bottom)
against $\omega t/\pi$ over one period
for the critically modulated temperature protocol~(\ref{strong})
at various reduced frequencies $\omega/a$ (see legend).
Black curves: predictions~(\ref{rclead}) of adiabatic approximation.
Black horizontal lines: stationary high-frequency values.}
\label{rcs}
\end{center}
\end{figure}

Figure~\ref{rcs} shows the reduced quantities
$2\rho(t)/\mu_0$ (top) and $2/(\mu_0\chi(t))$ (bottom),
obtained by means of~(\ref{rcser}) and
plotted against $\omega t/\pi$ over one period in the stationary state.
Black curves show the predictions~(\ref{rclead}) of the adiabatic approximation,
whereas black horizontal lines show the
stationary values of the observables in the high-frequency regime,
obtained by replacing the periodic protocol $\mu^2(t)$ by its average over one period,
\beq
\frac{\omega}{2\pi}\int_0^{2\pi/\omega}\mu^2(t)\,\dd t=a=\mu_0^2.
\eeq

The vanishing of the driving temperature at periodic instants of time
is expected to have its most drastic consequences at low frequency ($\omega\ll a$).
The form of $\mu(t)$ in the vicinity of each of these instants
is precisely of the form~(\ref{slomu}) of a single slow quench,
with the following parameter values
\beq
\alpha=2,\qquad\tau=\frac{1}{\omega},\qquad A=\frac{a}{2}.
\label{prams}
\eeq
The analysis of section~\ref{sloque} suggests
that the minima of the quantities plotted in figure~\ref{rcs}
take place near the instants where temperature vanishes, and scale as $\omega^{1/3}$.
This is in qualitative agreement with figure~\ref{rcs}.

\begin{figure}
\begin{center}
\includegraphics[angle=0,width=.7\linewidth,clip=true]{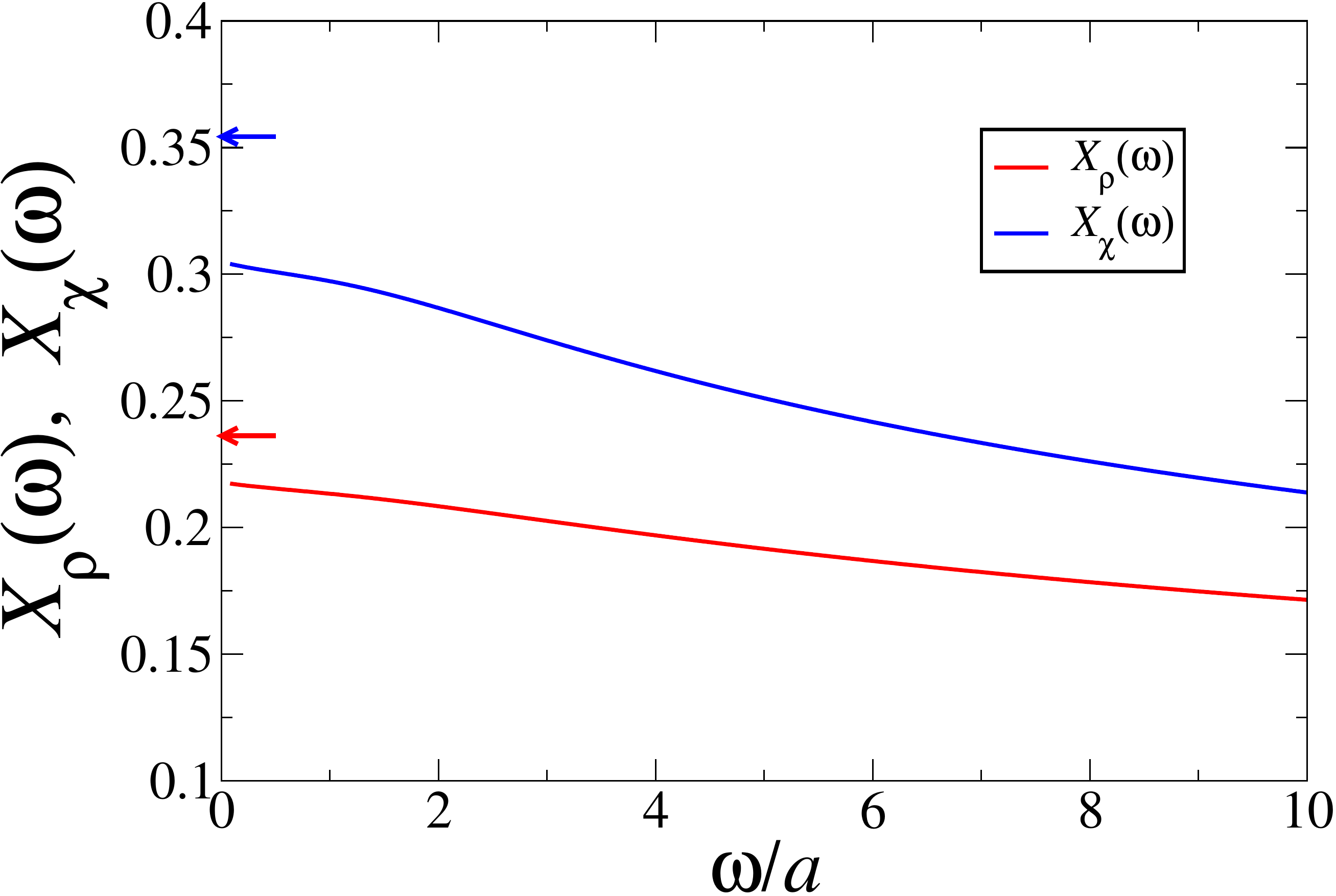}
\caption{\small
Reduced quantities $X_\rho(\omega)$ and $X_\chi(\omega)$ (see~(\ref{xdef}))
against reduced frequency $\omega/a$.
Arrows: exact limits $L_\rho$ and $L_\chi$ (see~(\ref{ldef}))
for single slow quenches with the same parameter values~(\ref{prams}).}
\label{mimirc}
\end{center}
\end{figure}

Let us now turn to a quantitative analysis.
Figure~\ref{mimirc} shows the reduced quantities
\beq
X_\rho(\omega)=\frac{\rho_\min(\omega)}{(a\omega^2)^{1/6}},\qquad
X_\chi(\omega)=\frac{1}{\chi_\max(\omega)\,(a\omega^2)^{1/6}},
\label{xdef}
\eeq
plotted against reduced frequency $\omega/a$,
where $\rho_\min(\omega)$ and $\chi_\max(\omega)$ are respectively the smallest
value of $\rho(t)$
and the largest value of $\chi(t)$ over one period in the stationary state at
frequency $\omega$.
The plotted quantities converge to the limiting values
$X_\rho(0)\approx0.218$ and $X_\chi(0)\approx0.304$.
These limits are some 8 and 14 percent smaller than their
counterparts for single slow quenches with the same parameter values~(\ref{prams}) (arrows),
namely (see~(\ref{rslo}),~(\ref{cslo}))
\beqa
L_\rho=\frac{\Gamma(5/6)}{2\,6^{1/6}\sqrt{\pi}}\approx0.236219,
\nonumber\\
L_\chi=\frac{\sqrt{\pi}}{4\,6^{1/6}\Gamma(7/6)}\approx0.354328.
\label{ldef}
\eeqa
We have $L_\rho/L_\chi=2/3$, in agreement with the limiting value $\Pi=2/3$
of the form factor (see~(\ref{pislo})).
The minima of the plots in figure~\ref{rcs}
exhibit some delay with respect to the instant where temperature vanishes,
and the limiting values $X_\rho(0)$ and~$X_\chi(0)$ are slightly smaller than their
counterparts~(\ref{ldef}) for slow quenches.
Both observations reflect that the system still cools and coarsens for a short while
after the instant where temperature vanishes.
A remarkable feature of the reduced quantities $X_\rho(\omega)$ and
$X_\chi(\omega)$
is their weak dependence on $\omega$ over a wide frequency range.

\section{Kovacs effect}
\label{ko}

The last protocol to be addressed in this work is the two-step protocol
leading to the memory effect~\cite{kov1,kov2} found by Kovacs in temperature
shift experiments on glassy polymers, where the volume (or energy) displays a
non-monotonic time behaviour.

\begin{figure}
\begin{center}
\includegraphics[angle=0,width=.6\linewidth,clip=true]{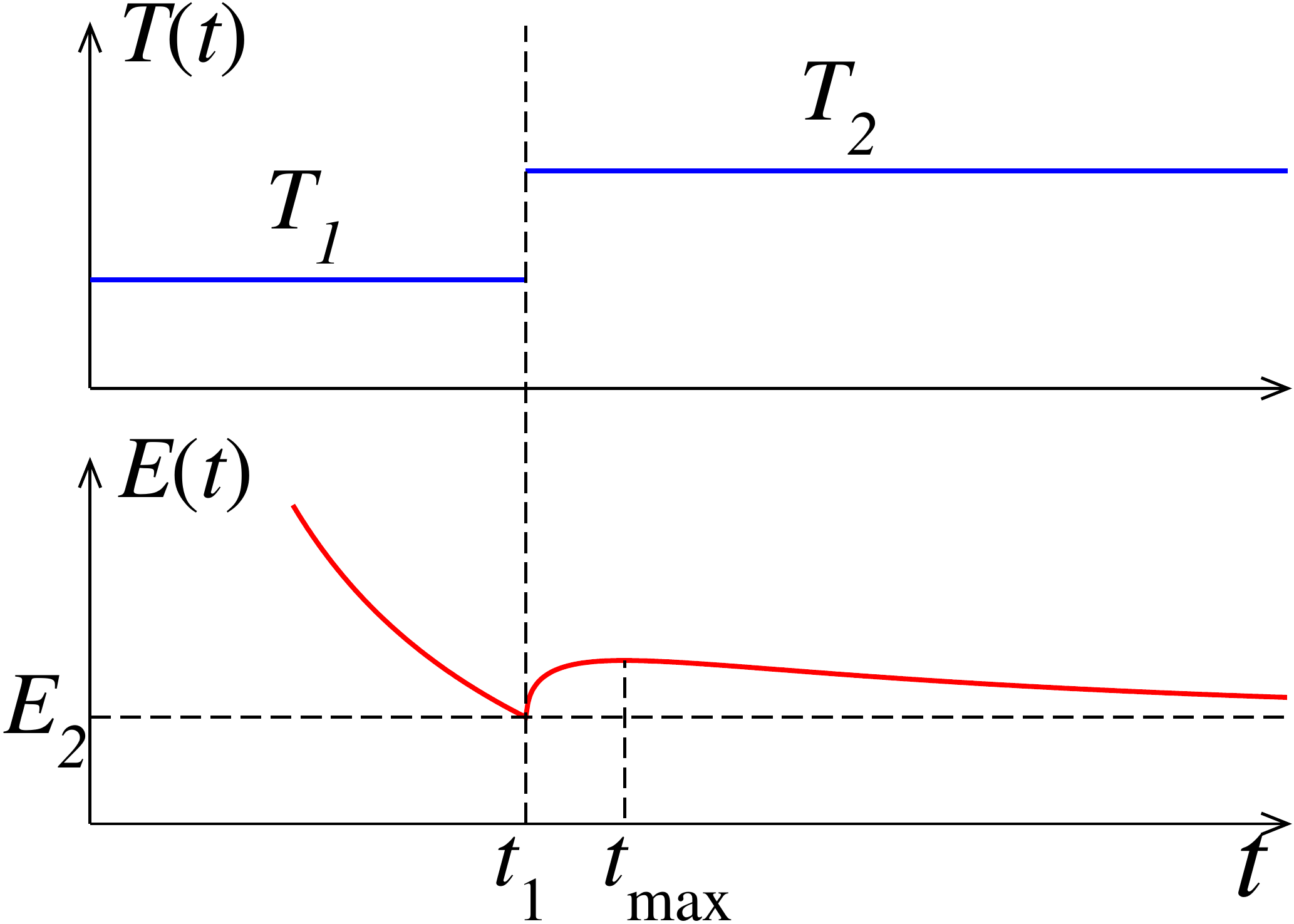}
\caption{\small
Sketchy plot of the two-temperature protocol leading to the Kovacs effect.}
\label{kovacs}
\end{center}
\end{figure}

The Kovacs protocol is sketched in figure~\ref{kovacs}.
The system is quenched from an equilibrium state at some initial temperature $T_0$
to a constant temperature~$T_1$ during a certain time $t_1$.
The temperature is then instantly raised from $T_1$ to $T_2$
(see the upper panel of figure~\ref{kovacs}).
The energy density $E(t)$ is monitored throughout the protocol.
The time $t_1$ is chosen such that the energy $E(t_1)$ is equal to
the equilibrium energy $E_2=E_\eq(T_2)$ at the final temperature $T_2$.
The Kovacs effect resides in the observation that the energy $E(t)$
at subsequent times ($t>t_1$) does not stay equal to $E_2$,
but rather exhibits a non-monotonic behaviour,
sketched in the lower panel of figure~\ref{kovacs},
first increasing up to some time $t_\max$
and then relaxing to its equilibrium value $E_2$,~i.e.,
\beq
E(t)=E_2+\Delta E(t),
\eeq
where $\Delta E(t)$ is dubbed the \textit{Kovacs hump}.

This well-documented feature of glassy dynamics
takes place in a broad class of systems,
including models without quenched disorder undergoing coarsening~\cite{bebo,bbd}.
The Kovacs effect in the Glauber-Ising chain has been addressed in two earlier works,
concerning respectively one specific situation~\cite{bra}
and a thorough analysis in the linear-response regime,
where the three temperatures~$T_0$,~$T_1$ and $T_2$ are close to each other~\cite{rgp}.
Our purpose is to provide a quantitative analysis of the Kovacs effect
throughout the low-temperature scaling regime,
where the Ising chain is launched from a random initial condition $(T_0\to\infty$)
and subjected to two low temperatures corresponding to the parameters $\mu_1$ and $\mu_2$.
This protocol is clearly far from the linear-response regime.
No direct comparison of the present results with those of~\cite{rgp} is therefore possible.

In the first stage ($t<t_1$),
the temperature parameter $\mu_1$ is constant,
so that the defect density $\rho(t)$ is given by~(\ref{rhosca}),
up to the replacement of $\mu$ by $\mu_1$, i.e.,
\beq
\rho(t)\approx\frac{\e^{-\mu_1^2t}}{2\sqrt{\pi
t}}+\frac{\mu_1}{2}\erf(\mu_1\sqrt{t}).
\label{rho1}
\eeq
In particular, the condition
\beq
\rho(t_1)=\rho_\eq(\mu_2)\approx\frac{\mu_2}{2}
\eeq
yields
\beq
\frac{\e^{-\mu_1^2t_1}}{\sqrt{\pi t_1}}+\mu_1\erf(\mu_1\sqrt{t_1})=\mu_2.
\label{t1cd}
\eeq
The above equation defines $t_1$ as a function of both temperature parameters
$\mu_1$ and~$\mu_2$.
Scaling implies that the combination $\mu_2^2t_1$ only depends on the ratio
\beq
q=\frac{\mu_1}{\mu_2}<1.
\label{qdef}
\eeq

In the second stage ($t>t_1$),
the temperature parameter $\mu_2$ is again constant,
whereas we have (see~(\ref{mdef}))
\beq
M(t)=\mu_1^2t_1+\mu_2^2(t-t_1).
\eeq
Inserting this expression into~(\ref{rhopro}) and performing the integral,
we are left with
\beqa
\rho(t)&\approx&\frac{\e^{-\mu_1^2t_1-\mu_2^2(t-t_1)}}{2\sqrt{\pi t}}
+\frac{\mu_2}{2}\erf(\mu_2\sqrt{t-t_1})
\nonumber\\
&+&\frac{\mu_1}{2}\e^{-(\mu_2^2-\mu_1^2)(t-t_1)}
\left(\erf(\mu_1\sqrt{t})-\erf(\mu_1\sqrt{t-t_1})\right).
\label{rho2}
\eeqa

\begin{figure}
\begin{center}
\includegraphics[angle=0,width=.7\linewidth,clip=true]{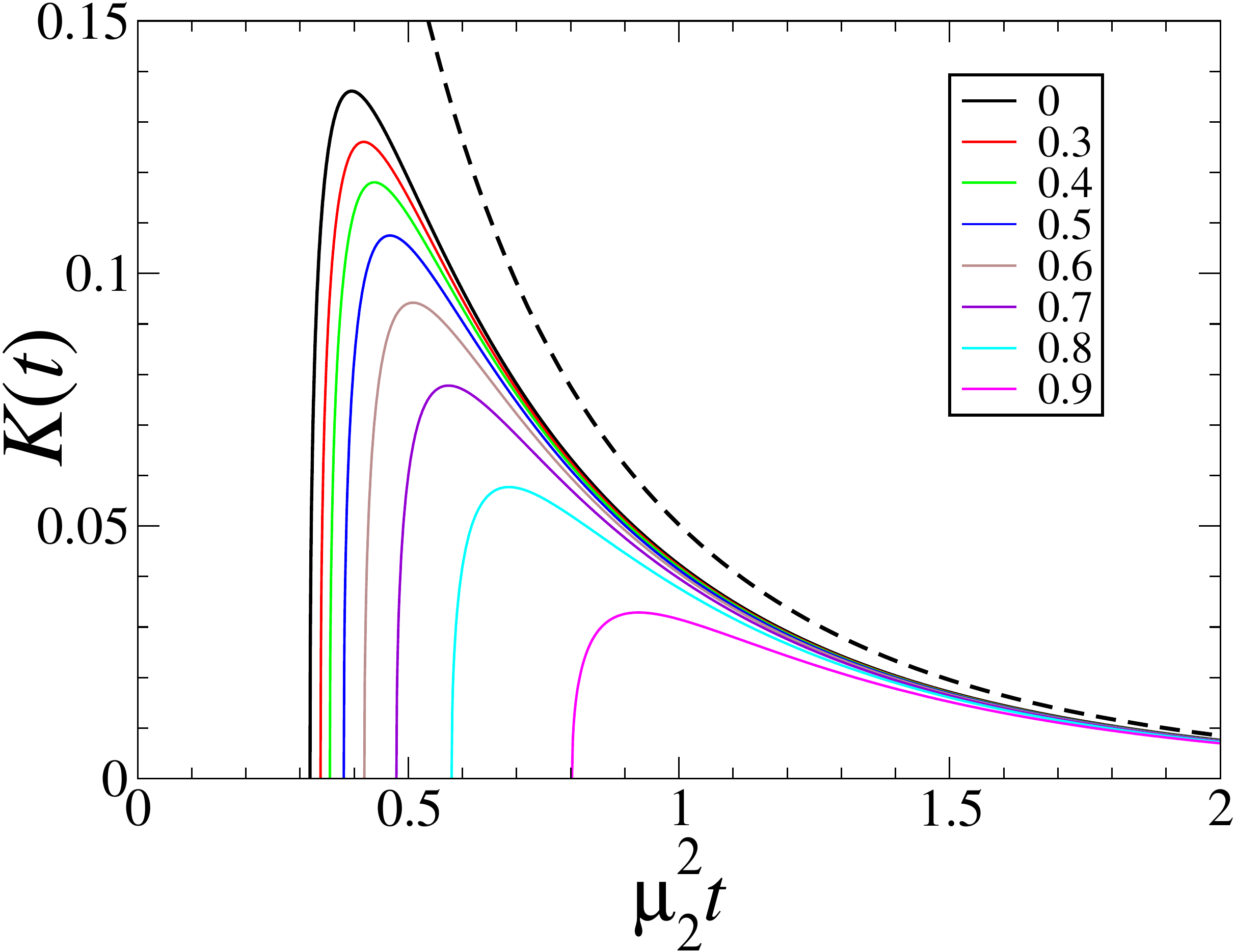}
\caption{\small
Full curves: dimensionless Kovacs hump function $K(t)$ (see~(\ref{kt}))
against reduced time $\mu_2^2t$
for several values of the ratio $q$ (see legend).
Dashed curve: dimensionless function $K_0(t)$ (see~(\ref{k0t}))
corresponding to an instantaneous quench to the final temperature.}
\label{kplot}
\end{center}
\end{figure}

The result~(\ref{rho2}) can be recast as
\beq
\rho(t)\approx\frac{\mu_2}{2}\,(1+K(t)),
\label{kdef}
\eeq
where we have introduced the dimensionless Kovacs hump function
\beqa
K(t)&=&\frac{\e^{-\mu_1^2t_1-\mu_2^2(t-t_1)}}{\mu_2\sqrt{\pi t}}
-\erfc(\mu_2\sqrt{t-t_1})
\nonumber\\
&+&\frac{\mu_1}{\mu_2}\e^{-(\mu_2^2-\mu_1^2)(t-t_1)}
\left(\erf(\mu_1\sqrt{t})-\erf(\mu_1\sqrt{t-t_1})\right).
\label{kt}
\eeqa
It is interesting to compare the hump function $K(t)$
to its counterpart $K_0(t)$
in the case of an instantaneous quench to the final temperature parameter $\mu_2$.
Setting $t_1=0$ in~(\ref{kt}), we obtain
\beq
K_0(t)\approx\frac{\e^{-\mu_2^2t}}{\mu_2\sqrt{\pi t}}-\erfc(\mu_2\sqrt{t}),
\label{k0t}
\eeq
in full agreement with~(\ref{rhosca}).
Figure~\ref{kplot} shows a comparison between
the hump function~$K(t)$ for several values of the ratio $q$
and its counterpart $K_0(t)$.

The hump function starts from the value $K(t_1)=0$, as should be.
The initial rise
\beq
K(t)\approx\frac{2}{\mu_2\sqrt{\pi}}(\mu_2^2-\mu_1^2)\sqrt{t-t_1}
\label{kini}
\eeq
of the expression~(\ref{kt}) has to be taken with a grain of salt.
For any finite temperatures,~$K(t)$ starts rising in a steep but regular way
over a microscopic time interval after $t_1$.
The law~(\ref{kini}) holds for $t-t_1$ much larger than this microscopic time,
but much smaller than $1/\mu_2^2$.
This subtlety of the continuum framework derived in section~\ref{low}
and used throughout this paper only shows up here,
because the Kovacs protocol $\mu(t)$ is discontinuous at $t_1$.

\begin{figure}
\begin{center}
\vskip 10pt
\includegraphics[angle=0,width=.7\linewidth,clip=true]{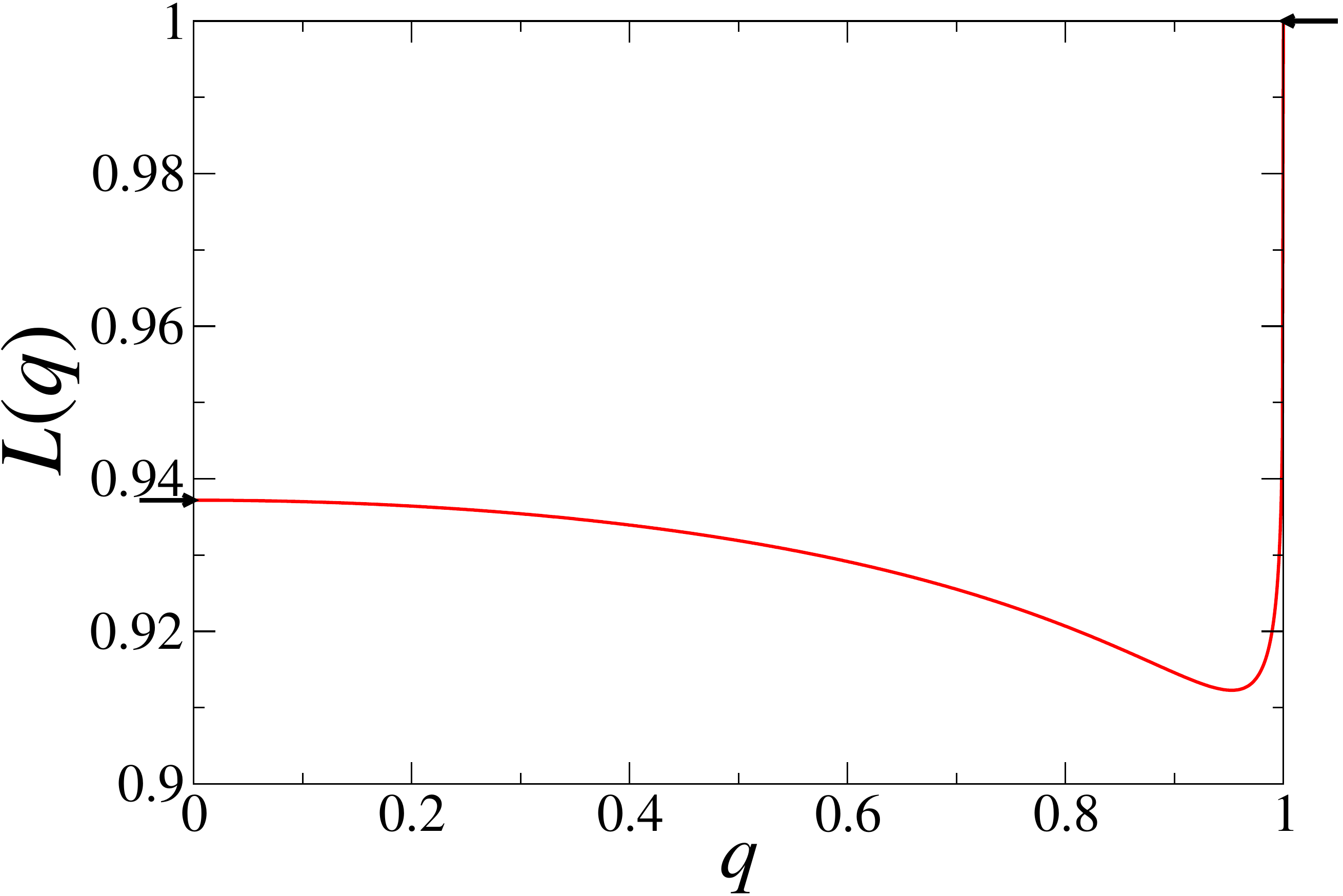}
\caption{\small
Kovacs effect:
limit ratio $L(q)$ against ratio $q$.
Arrows: limiting values~(\ref{lqz}) for $q=0$ and 1 for $q=1$
(see~(\ref{lqlog})).}
\label{lq}
\end{center}
\end{figure}

The hump function $K(t)$ falls off exponentially at large times, according~to
\beq
K(t)\approx\frac{\e^{-\mu_2^2(t-t_1)}}{2\mu_2\sqrt{\pi t^3}}
\left(\frac{1}{\mu_2^2}-\frac{1-\e^{-\mu_1^2t}}{\mu_1^2}\right).
\eeq
Both $K(t)$ and $K_0(t)$ have the same decay law at large times.
The limit ratio
\beq
L(q)=\lim_{t\to\infty}\frac{K(t)}{K_0(t)}
\eeq
only depends on the temperature ratio $q$, and reads explicitly
\beq
L(q)=\e^{\mu_2^2t_1}\left(1-\frac{1-\e^{-q^2\mu_2^2t_1}}{q^2}\right).
\eeq
This quantity is plotted against $q$ in figure~\ref{lq}.
It is slightly smaller than unity for all values of $q$,
so that the tail of the hump function $K(t)$ is slightly below
that of its counterpart~$K_0(t)$, as can be seen in figure~\ref{kplot}.
The limit ratio $L(q)$ exhibits a non-monotonic dependence on $q$,
starting from the value~(\ref{lqz}) at $q=0$,
reaching its minimum $L_\min\approx0.912273$ for $q\approx0.95254$,
and converging very steeply to unity in a tiny vicinity of $q=1$,
according to~(\ref{lqlog}).

More specific results can be obtained in the limiting situations $q=0$ and $q\to1$.
For $\mu_1=0$ ($q=0$),
i.e., when the first cooling temperature is exactly zero,
the Kovacs hump is the largest (see black curve in figure~\ref{kplot}).
We have then
\beq
\mu_2^2t_1\approx\frac{1}{\pi}\approx0.318309,
\label{t10}
\eeq
and so
\beq
L(0)=\e^{1/\pi}(1-1/\pi)\approx0.937189,
\label{lqz}
\eeq
whereas~(\ref{kt}) simplifies to
\beq
K(t)\approx\frac{\e^{-(\mu_2^2t-1/\pi)}}{\mu_2\sqrt{\pi t}}-\erfc\sqrt{\mu_2^2t-1/\pi}.
\eeq
The maximum of the Kovacs hump,
\beq
K_\max\approx0.136094,
\label{k0}
\eeq
is reached for
\beq
\mu_2^2t_\max\approx0.395423.
\label{tmax0}
\eeq

In the other limiting situation, i.e., $\mu_1\to\mu_2$ or $q\to1$,
the time $t_1$ diverges logarithmically, according to
\beq
\mu_2^2t_1\approx\abs{\ln(1-q)}-\half\ln(4\pi\abs{\ln(1-q)}^3),
\label{t11}
\eeq
and so
\beq
L(q)\approx1-\frac{1}{\sqrt{\pi\abs{\ln(1-q)}^3}},
\label{lqlog}
\eeq
whereas the whole Kovacs hump vanishes proportionally to $1-q$.
Equation~(\ref{kt}) indeed simplifies to
\beq
K(t)\approx(1-q)\phi(z),
\label{kphi}
\eeq
with
\beqa
z=\mu_2^2(t-t_1),
\nonumber\\
\phi(z)=\left(2\sqrt\frac{z}{\pi}+1\right)\e^{-z}-(2z+1)\erfc\sqrt{z}.
\eeqa
The maximum of the scaling function,
\beq
\phi_\max\approx0.499014,
\eeq
so that
\beq
K_\max\approx(1-q)\phi_\max,
\label{k1}
\eeq
is reached for
\beq
z_\max\approx0.257458.
\label{zmax}
\eeq

\begin{figure}
\begin{center}
\includegraphics[angle=0,width=.7\linewidth,clip=true]{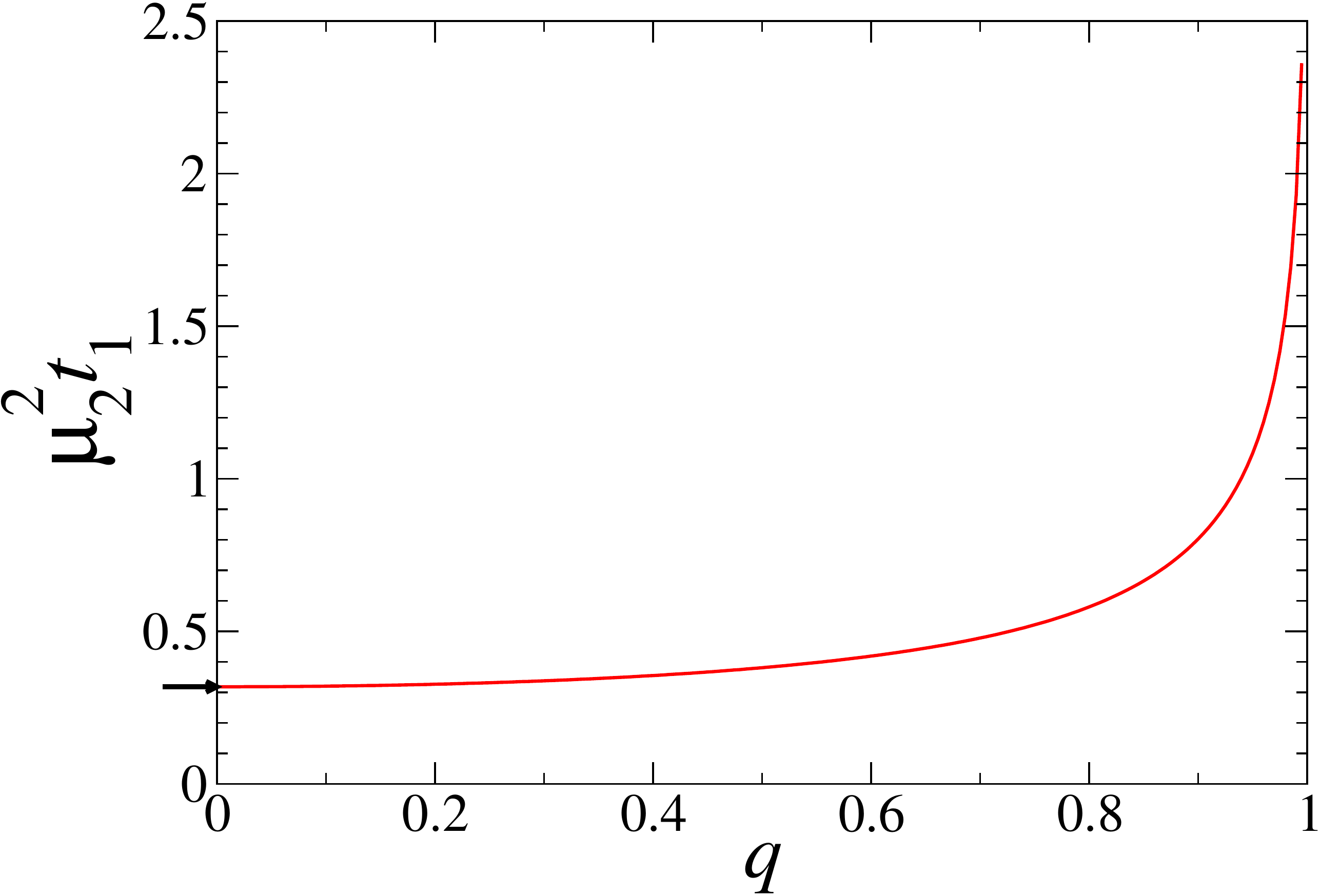}
\caption{\small
Kovacs effect:
combination $\mu_2^2t_1$ against ratio $q$.
Arrow: minimal value~(\ref{t10}).}
\label{tone}
\end{center}
\end{figure}

\begin{figure}
\begin{center}
\includegraphics[angle=0,width=.7\linewidth,clip=true]{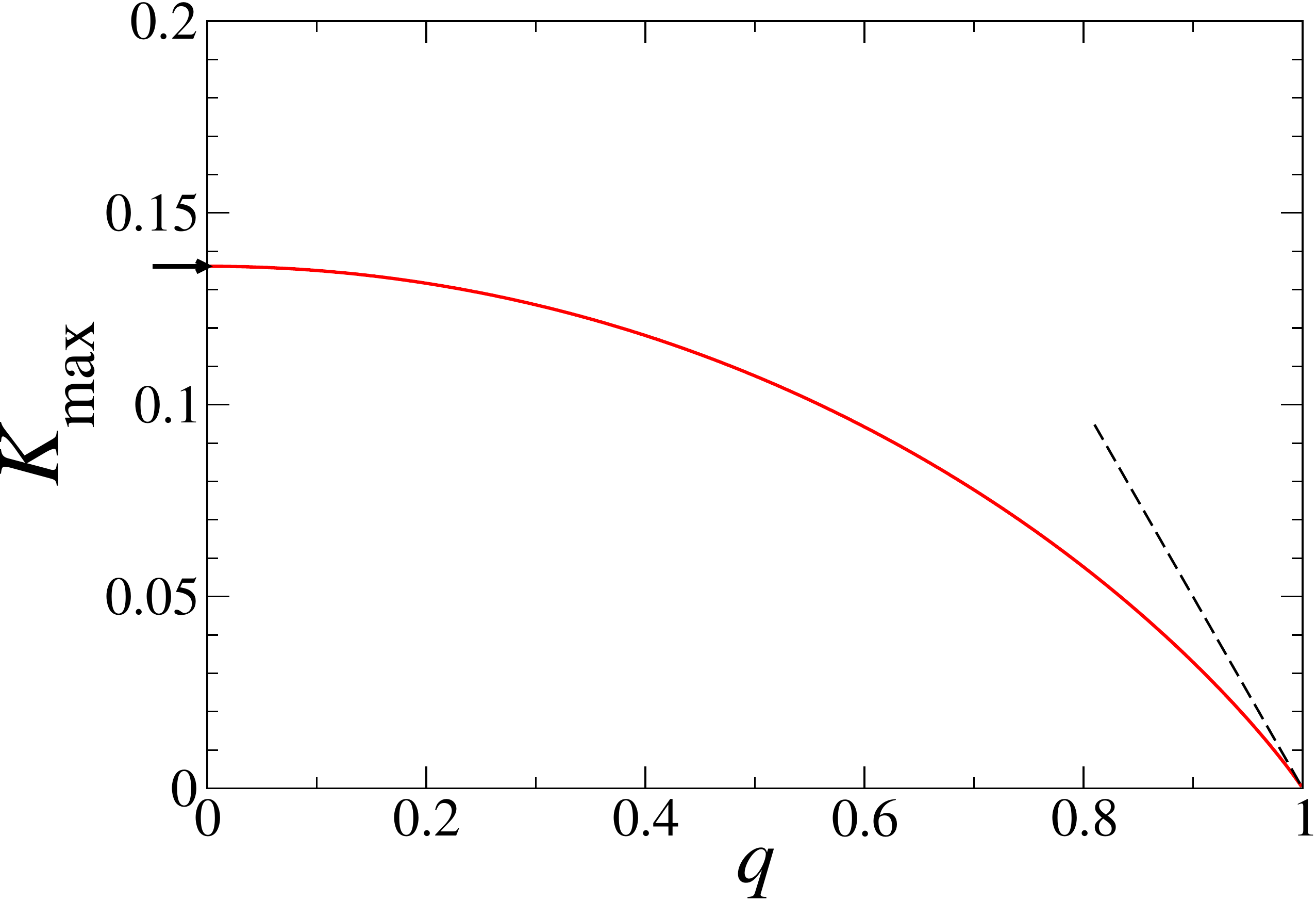}
\caption{\small
Kovacs effect:
maximum $K_\max$ of hump function $K(t)$ against ratio~$q$.
Arrow: maximal value~(\ref{k0}).
Dashed line: linear law~(\ref{k1}).}
\label{kmax}
\end{center}
\end{figure}

\begin{figure}
\begin{center}
\vskip 10pt
\includegraphics[angle=0,width=.7\linewidth,clip=true]{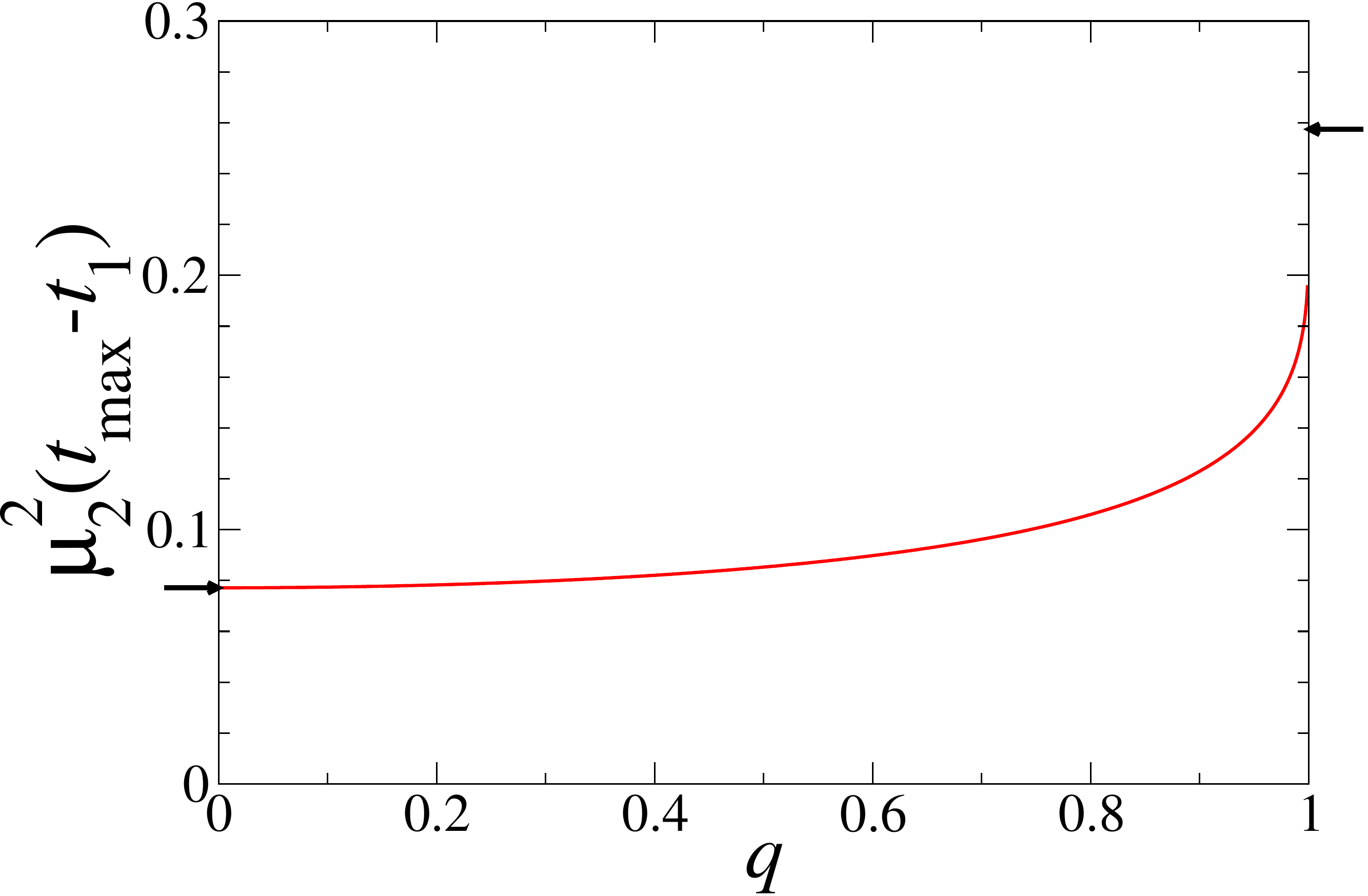}
\caption{\small
Kovacs effect:
combination $\mu_2^2(t_\max-t_1)$ against ratio~$q$.
Arrows: limiting values 0.077113 for $q=0$ (see~(\ref{t10}),~(\ref{tmax0}))
and $z_\max$ for $q=1$ (see~(\ref{zmax})).}
\label{tmax}
\end{center}
\end{figure}

Figure~\ref{kplot} demonstrates that the Kovacs hump shifts to the right,
whereas its amplitude decreases, as the ratio $q$ increases from 0 to 1.
This trend can be monitored by means of several quantities.
Figure~\ref{tone} shows the combination~$\mu_2^2t_1$ against the ratio~$q$.
This quantity increases with~$q$,
starting from the value~(\ref{t10}) for $q=0$ (arrow).
The logarithmic law~(\ref{t11}) only holds in a tiny vicinity of $q=1$.
Figure~\ref{kmax} shows the maximum $K_\max$ of the Kovacs hump against $q$.
This quantity decreases with~$q$,
starting from the value~(\ref{k0}) for $q=0$ (arrow).
The linear law~(\ref{k1}) as $q\to1$ (dashed line)
is affected by strong subleading corrections.
Finally, figure~\ref{tmax} shows the combination $\mu_2^2(t_\max-t_1)$ against~$q$.
This quantity increases monotonically with~$q$
between the limiting values (arrows)
0.077113 for $q=0$ (see~(\ref{t10}),~(\ref{tmax0}))
and~$z_\max$ for $q=1$ (see~(\ref{zmax})).
The second limit is reached very steeply in a tiny vicinity of $q=1$.
This is again due to the presence of strong subleading corrections
to the estimates~(\ref{t11}) and~(\ref{kphi}).

\section{Discussion}
\label{disc}

As recalled in the introduction of this paper,
there has been a continuous effort
aimed at extending in various directions the pioneering work of
Glauber~\cite{glauber}
on the exactly solvable single-spin-flip dynamics of the ferromagnetic Ising chain.
A first series of papers were devoted to the general analysis
of arbitrary temperature protocols~\cite{reiss,schilling,brey}.
Then, more recently, slow quenches have been considered
and put in perspective with the Kibble-Zurek scaling
theory~\cite{krapiv,jeong,priyanka,mayo}.

The present work aims at revisiting and completing those earlier works in two
main directions.
First, we demonstrate that
focussing our attention on the low-temperature scaling regime
provides a considerable simplification of earlier treatments.
Second, we consider in parallel two concurrent observables
probing the two-point correlation function at two very different spatial scales,
namely the density $\rho(t)$ of domain walls and the reduced susceptibility $\chi(t)$.
The central outcome of this approach
resides in the expressions~(\ref{rhopro}) and~(\ref{chipro}),
giving closed-form integral representations of both key observables
for an arbitrary protocol in the low-temperature regime.
The dimensionless form factor $\Pi(t)=\rho(t)\chi(t)$
characterises the pattern of growing ordered domains by a single number
and provides a measure of the distance of the system to thermal equilibrium.

This scheme is used in sections~\ref{ever} to~\ref{ko}
to investigate thoroughly a range of protocols in the low-temperature scaling regime.
Among the main features of these specific outcomes,
we would like to emphasise once more the importance of the form factor~$\Pi(t)$
and to underline its rich dependence on model parameters.
As announced in the introduction,
the non-triviality of $\Pi(t)$ mirrors the property that domain lengths
are neither mutually independent nor exponentially distributed~\cite{dz},
except at thermal equilibrium, where we have $\Pi_\eq=1$.
The universal value $\Pi_\coa=2/\pi$ in the zero-temperature coarsening regime
is very likely to provide an absolute lower bound for~$\Pi(t)$
in the low-temperature scaling regime, irrespective of the driving protocol.
The form factor exhibits an interesting dependence on model parameters
in several different situations.
It departs from its equilibrium value in a controlled way
for generic slowly varying temperature protocols (see~(\ref{fulltauslow})).
It depends continuously on $\mu^2t\approx t/\tau_\eq$ (see~(\ref{pisca}) and
figure~\ref{pilow})
for quenches to a constant low temperature,
on the amplitude $B$ (see~(\ref{pib})) of critical everlasting quenches,
and on the exponent~$\alpha$ (i.e., on the Kibble-Zurek exponent $\lambda$)
(see~(\ref{pislo})) right at the end of slow quenches of finite duration.

To close, let us mention that considering more complex correlation and response functions,
involving four or a higher even number of spins, possibly at different times,
somewhat along the lines of~\cite{MS},
may certainly reveal yet other features of interest
in the Glauber-Ising chain under diverse low-temperature protocols.
In this respect,
it is worth underlining that the fermionic functional formalism
introduced by Aliev in the framework of the dynamics at constant
temperature~\cite{ali1},
which provides an efficient compact way of dealing with all equal-time
correlations,
has been shown to hold in the presence of quenched disorder~\cite{ali3}
and to extend to arbitrary time-dependent temperature protocols~\cite{aliev}.

\ack

It is a pleasure to thank Paul Krapivsky for fruitful discussions.
Part of the results obtained in this paper were presented in a talk by JML
at the conference {\it Statistical Physics and Low Dimensional Systems}
honoring Malte Henkel on the occasion of his 60th birthday.

\appendix
\section{Fourier and Laplace transforms}
\label{appfl}

We begin by recalling the definitions of the Laplace and (discrete and continuous)
Fourier transforms that are used in the body of this paper.

\begin{itemize}

\item
Laplace transform of a function $f(t)$ $(t\ge0)$:
\beq
f^\L(p)=\int_0^\infty\e^{-pt}\,f(t)\dd t,
\quad
f(t)=\int_{c-\ii\infty}^{c+\ii\infty}\frac{\dd p}{2\pi\ii}\,\e^{pt}\,f^\L(p)
\eeq
for some $c>0$.

\item
Discrete Fourier transform of a function $f_n$ defined on the lattice of integers:
\beq
f^\F(q)=\sum_{n=-\infty}^{+\infty}\e^{-\ii nq}\,f_n,
\quad
f_n=\int_0^{2\pi}\frac{\dd q}{2\pi}\,\e^{\ii nq}\,f^\F(q).
\eeq
Alternatively,
up to a change of the sign of the integer $n$,
the $f_n$ are the coefficients of the Fourier series
representing the $2\pi$-periodic function $f^\F(q)$.

\item
Continuous Fourier transform of a function $f(x)$ defined on the real line:
\beq
f^\F(q)=\int_{-\infty}^{+\infty}\e^{-\ii qx}\,f(x)\dd x,
\quad
f(x)=\int_{-\infty}^{+\infty}\frac{\dd q}{2\pi}\,\e^{\ii qx}\,f^\F(q).
\eeq

\end{itemize}

We give in table~\ref{ilts} a few inverse Laplace transforms
involving modified Bessel and error functions, which are used in the body of this article.

\begin{table}
\begin{center}
$$
\begin{array}{|l|l|l|}
\hline
& f^\L(p) & f(t)\\
\hline
(1)
& \frac{1}{\sqrt{p^2-a^2}} & I_0(at)\\
(2)
& \sqrt\frac{p+a+b}{p+a-b}-1 & b\e^{-at}\left(I_0(bt)+I_1(bt)\right)\\
(3)
& \frac{2}{p}\,\e^{-b\sqrt{p+a^2}}
& \e^{ab}\,\erfc\frac{b+2at}{2\sqrt{t}}+\e^{-ab}\,\erfc\frac{b-2at}{2\sqrt{t}}\\
(4)
& \frac{\sqrt{p+a^2}}{p} & \frac{\e^{-a^2t}}{\sqrt{\pi t}}+a\,\erf(a\sqrt{t})\\
(5)
& \frac{a}{p\sqrt{p+a^2}} & \erf(a\sqrt{t})\\
\hline
\end{array}
$$
\caption
{A few useful inverse Laplace transforms involving modified Bessel and error functions.
The symbols $a$, $b$ denote positive real numbers.}
\label{ilts}
\end{center}
\end{table}

\section{Modified Bessel functions}
\label{appb}

We recall below a few useful properties of the modified Bessel functions.
The modified Bessel functions of integer order $n$ are defined for complex $z$
as the integrals
\beq
I_n(z)=\int_0^{2\pi}\frac{\dd\theta}{2\pi}\,\e^{-\ii n\theta+z\cos\theta}.
\label{bdef}
\eeq
They are analytic functions in the entire $z$ plane.
Their power-series expansions read
\beq
I_n(z)=\sum_{k\ge0}\frac{(z/2)^{n+2k}}{k!(n+k)!}\qquad(n\ge0).
\label{bser}
\eeq
They obey the symmetries
\beq
I_{-n}(z)=I_n(z),\qquad I_n(-z)=(-)^nI_n(z)
\label{bsym}
\eeq
and the difference equation
\beq
I_{n-1}(z)-I_{n+1}(z)=\frac{2n}{z}\,I_n(z).
\label{bdiff}
\eeq

\section*{References}

\bibliography{paper.bib}

\providecommand{\newblock}{}
\begin{thebibliography}{10}
\expandafter\ifx\csname url\endcsname\relax
  \def\url#1{{\tt #1}}\fi
\expandafter\ifx\csname urlprefix\endcsname\relax\def\urlprefix{URL }\fi
\providecommand{\eprint}[2][]{\url{#2}}

\bibitem{glauber}
Glauber R~J 1963 {\em J. Math. Phys.\/} {\bf 4} 294--307

\bibitem{reiss}
Reiss H 1980 {\em Chem. Phys.\/} {\bf 47} 15--24

\bibitem{schilling}
Schilling R 1988 {\em J. Stat. Phys.\/} {\bf 53} 1227--1235

\bibitem{brey}
Brey J~J and Prados A 1994 {\em Phys. Rev. B\/} {\bf 49} 984--997

\bibitem{kibble}
Kibble T~W~B 1980 {\em Phys. Rep.\/} {\bf 67} 183--199

\bibitem{zurek}
Zurek W~H 1996 {\em Phys. Rep.\/} {\bf 276} 177--221

\bibitem{krapiv}
Krapivsky P~L 2010 {\em J. Stat. Mech.\/} {\bf P02014}

\bibitem{jeong}
Jeong K, Kim B and Lee S~J 2020 {\em Phys. Rev. E\/} {\bf 102} 012114

\bibitem{priyanka}
Priyanka, Chatterjee S and Jain K 2021 {\em J. Stat. Mech.\/} {\bf 033208}

\bibitem{mayo}
Mayo J~J, Fan Z, Chern G~W and \protect{del Campo} A 2021 {\em Phys. Rev.
  Research\/} {\bf 3} 033150

\bibitem{biroli}
Biroli G, Cugliandolo L~F and Sicilia A 2010 {\em Phys. Rev. E\/} {\bf 81}
  050101

\bibitem{ceg}
Chandran A, Erez A, Gubser S~S and Sondhi S~L 2012 {\em Phys. Rev. B\/} {\bf
  86} 064304

\bibitem{dz}
Derrida B and Zeitak R 1996 {\em Phys. Rev. E\/} {\bf 54} 2513--2525

\bibitem{gl2000}
Godr\`eche C and Luck J 2000 {\em J. Phys. A: Math. Gen.\/} {\bf 33} 1151--1169

\bibitem{baxter}
Baxter R~J 1989 {\em Exactly Solved Models in Statistical Mechanics\/} (London:
  Academic Press)

\bibitem{bray}
Bray A~J 1994 {\em Adv. Phys.\/} {\bf 43} 357--459

\bibitem{wallis}
Chen C~P and Paris R~B 2017 {\em Applied Math. Comput.\/} {\bf 293} 30--39

\bibitem{YB}
Yi S~D and Baek S~K 2015 {\em Phys. Rev. E\/} {\bf 91} 062107

\bibitem{kov1}
Kovacs A~J 1963 {\em Fortschr. Hochpolym. Forsch. -- Adv. Polym. Sci.\/} {\bf
  3} 394--507

\bibitem{kov2}
Kovacs A~J, Aklonis J~J, Hutchinson J~M and Ramos A~R 1979 {\em J. Polym. Sci.
  B\/} {\bf 17} 1097--1162

\bibitem{bebo}
Berthier L and Bouchaud J~P 2002 {\em Phys. Rev. B\/} {\bf 66} 054404

\bibitem{bbd}
Bertin E~M, Bouchaud J~P, Drouffe J~M and Godr\`eche C 2003 {\em J. Phys. A:
  Math. Gen.\/} {\bf 36} 10701--10719

\bibitem{bra}
Brawer S 1978 {\em Phys. Chem. Glasses\/} {\bf 19} 48--51

\bibitem{rgp}
Ruiz-Garcia M and Prados A 2014 {\em Phys. Rev. E\/} {\bf 89} 012140

\bibitem{MS}
Mayer P and Sollich P 2004 {\em J. Phys. A: Math. Gen.\/} {\bf 37} 9--49

\bibitem{ali1}
Aliev M~A 1998 {\em Phys. Lett. A\/} {\bf 241} 19--27

\bibitem{ali3}
Aliev M~A 2000 {\em Physica A\/} {\bf 277} 261--273

\bibitem{aliev}
Aliev M~A 2009 {\em J. Math. Phys.\/} {\bf 50} 083302

\end{thebibliography}

\end{document}